\documentstyle[graphicx]{mn}

\title{A \ro HRI survey of bright nearby galaxies}

\author[T.\,P. Roberts and R.\,S. Warwick ]
{T.\,P. Roberts and R.\,S. Warwick \\
Department of Physics and Astronomy, University 
of Leicester, University Road, Leicester, LE1 7RH}
\date{}

\def\ro{{\it ROSAT~\/}}

\def\ein{{\it EINSTEIN~\/}}
\def\xmm{{\it XMM~\/}}

\def\Lsun{\hbox{$\rm ~L_{\odot}$}}
\def\Msun{\hbox{$\rm ~M_{\odot}$}}

\def\ergsec{{\rm ~erg~s^{-1}}}

\def\atpcm{{\rm ~atoms~cm^{-2}}}
\def\ctsec{{\rm ~count~s^{-1}}}

\def\H0{{\rm ~km~s^{-1}~Mpc^{-1}}}

\def\eg{{\it e.g.~\/}}

\def\ie{{\it i.e.~\/}}

\def\la{\mathrel{\hbox{\rlap{\hbox{\lower4pt\hbox{$\sim$}}}{\raise2pt\hbox{$<$}}}}}
\def\ga{\mathrel{\hbox{\rlap{\hbox{\lower4pt\hbox{$\sim$}}}{\raise2pt\hbox{$>$}}}}}

\def\d25{D$_{25}$}
\def\nh{{$N_{\rm H}$}}
\def\Ha{{H$\alpha$}}

\def\hii{H {\small II}$~$}
\def\los{line-of-sight\thinspace}
\def\.25{0.25 keV\thinspace}
\def\lx{L$_{\rm X}$}

\begin{document}

\maketitle

\begin{abstract}

We use the extensive public archive of \ro High Resolution Imager
(HRI) observations to carry out a statistical investigation of the
X-ray properties of nearby galaxies. Specifically we focus on the
sample of 486 bright ($B_T \leq 12.5$) northern galaxies studied by
Ho, Filippenko and Sargent (HFS) in the context of their exploration
of the optical spectroscopic properties of nearby galactic
nuclei. Over $20\%$ of HFS galaxies are encompassed in \ro HRI fields
of reasonable ($ \geq 10$ks) exposure. The X-ray sources detected
within the optical extent of each galaxy are categorised as either
nuclear or non-nuclear depending on whether the source is positioned
within or outside of a $25''$ radius circle centred on the optical
nucleus.

A nuclear X-ray source is detected in over 70\% of the galaxies
harbouring either a Seyfert or LINER nucleus compared to a detection
rate of only $\sim 40\%$ in less active systems.  The correlation of
the \Ha~luminosity with nuclear X-ray luminosity previously observed
in QSOs and bright Seyfert 1 galaxies appears to extend down into the
regime of ultra-low luminosity (\lx $\sim 10^{38}-10^{40} \ergsec$)
active galactic nuclei (AGN). The inferred accretion rates for this
sample of low-luminosity AGN are significantly sub-Eddington.

In total 142 non-nuclear sources were detected. In combination with
published data for M31 this leads to a luminosity distribution
(normalised to an optical blue luminosity of $10^{10} \Lsun$) for the
discrete X-ray source population in spiral galaxies of the form
$dN/dL_{38} = (1.0 \pm 0.2) L_{38}^{-1.8}$, where $L_{38}$ is the
X-ray luminosity in units of $10^{38} \ergsec$. The implied
\lx/L$_{B}$ ratio is $\sim 1.1 \times 10^{39} \ergsec (10^{10}
\Lsun)^{-1}$. The nature of the substantial number of
``super-luminous'' non-nuclear objects detected in the survey is
discussed.

\end{abstract}


\section{Introduction}

The \ro public archive now includes over 4400 pointed observations
carried out with the \ro High Resolution Imager (HRI) during the
period June 1990 to December 1997.  One obvious application of this
extensive database is the investigation of the X-ray properties of
particular classes of object, especially those which are available in
large numbers as the result of detailed surveys carried out in other
wavebands. This is the approach adopted in the present paper which
focuses on the luminous discrete X-ray sources associated with nearby
bright galaxies.

For this purpose we use the sample of 486 bright ($B_T \leq 12.5$)
predominantly northern ($\delta > 0^\circ$) galaxies studied by Ho,
Filippenko \& Sargent (1995) in the context of their search for
``dwarf'' Seyfert nuclei in nearby galaxies.  It turns out that, of
the galaxies studied by Ho, Filippenko and Sargent (hereafter the HFS
sample), over $20\%$ lie within the field of view of reasonably deep
($ \geq 10$ks exposure) \ro HRI observations.  As part of their
programme Ho, Filippenko \& Sargent (1995, 1997a,b,c) have compiled a
huge database of information pertaining to both the optical
spectroscopic properties of the galaxy nuclei and the properties of
the host galaxy. This comprehensive and coherent body of information
provides an ideal resource for our present purpose.

The first detailed studies of the X-ray properties of nearby galaxies
utilised the X-ray imaging and spectroscopic capabilities of the
\ein Observatory (as summarised in the \ein results atlas
published by Fabbiano, Kim \& Trinchieri 1992 and more recently in the
work of Burstein et al. 1997). The Einstein observations demonstrated
that the X-ray emission from bright elliptical and lenticular galaxies
is often dominated by their hot interstellar medium, whereas in normal
spirals it is the integrated emission of evolved stellar sources, such
as supernova remnants (SNRs) and X-ray binaries (XRBs) which is
generally most important (see Fabbiano 1989 for a review). \ein
studies also established that, barring the presence of a luminous
active galactic nucleus (AGN), the X-ray luminosity exhibited by
normal galaxies is typically in the range $10^{38} - 10^{42}
\rm~erg~s^{-1}$. \ro observations have now added considerably to this
picture. Specifically the high sensitivity, soft bandpass, modest
spectral resolution and $< 30''$ spatial resolution of the \ro X-ray
telescope/position sensitive proportional counter (PSPC) combination
has greatly extended our knowledge of diffuse $10^{6} - 10^{7}$ K
thermal emission in galaxies and has confirmed the ubiquity of such
emission not only in early type and starburst galaxies but also as a
significant component in normal spirals (e.g. Read, Ponman \&
Strickland 1997).  Both \ro PSPC and HRI observations have also
furthered our knowledge of the resolved point X-ray source population
in nearby galaxies, for example with the detection of hundreds of such
sources in M31 (Primini et al. 1993; Supper et al. 1997).

Although the \ro PSPC provided somewhat more sensitivity than the \ro
HRI, the advantage of using data from the latter instrument is that,
by and large, it is possible to focus the investigation on a single
theme, namely the resolved discrete X-ray sources associated with
nearby galaxies.  Moreover the high spatial resolution afforded by the
\ro HRI (FWHM $<10''$) is particularly relevant when attempting to
distinguish the emission of putative low-luminosity AGN from other
luminous source populations in galaxies. The remainder of this paper
is ordered as follows. In section 2 we describe the available database
of HRI observations, the possible impact of selection effects in the
X-ray data and the method used to detect point X-ray sources in each
HRI field. Next, in section 3, we carry out a preliminary analysis of
the X-ray source sample, distinguishing between the ``nuclear'' and
``non-nuclear'' source populations. We also estimate the likely level
of contamination by foreground stars and background QSOs. Section 4
then investigates the incidence of luminous nuclear X-ray sources in
galaxies with nuclei of different optical spectroscopic type. Section
5 goes on to consider the X-ray luminosity distribution of the
non-nuclear source population and the nature of the substantial subset
of ``super-luminous'' sources. Finally in section 6 we provide a brief
summary of our findings.

\section{The X-ray database}

The relevant subset of \ro HRI pointed observations was identified by
cross-correlating the ROSPUBLIC database (circa July 1998) available
from the Leicester Data Archive Service (LEDAS) with the full list of
HFS galaxy ({\it i.e.} optical nuclear) positions.  All HRI fields
which encompassed the HFS galaxy position within $15'$ of the pointing
position and had an exposure of at least 10 ks were initially
accepted.  This process led to a set of 148 HRI observations of 106
HFS galaxies (with many of the galaxies having been observed more than
once).  However, a number of these observations were unsuitable for
our purposes. The two dominant Local Group members, NGC 224 (M 31) and
NGC 598 (M 33), were not included as their full angular extent cannot
be imaged in a single HRI observation. The HRI observations of NGC
1275 and NGC 4478/4486 were not used since these galaxies lie at the
centre of galaxy clusters (Perseus and Virgo respectively) and the
presence of a bright intra-cluster medium confuses the X-ray field and
limits the point source sensitivity.  Similarly to avoid the
complication of the dominant hot interstellar medium in massive early
type galaxies, ellipticals and lenticulars with $M_B < -20.5$ were
excluded, as was the bright starburst galaxy NGC 3034 (M82). Finally
the galaxies NGC 1052 and 4594 were rejected as southern interlopers
in the HFS sample.  A total of 83 galaxies remained, covered by 111
HRI observations, representing roughly a $17\%$ sampling of the full
HFS catalogue. Hereafter we refer to our HRI-observed subset of HFS
galaxies as the XHFS sample.

The parent HFS sample is reasonably complete in statistical terms (see
Ho, Filippenko \& Sargent 1995) and can certainly be considered as
representative of the nearby bright galaxy population. However, our
XHFS sample is subject to additional selection biases arising both
from the selection criteria noted above and from the fact that quite a
high fraction of the galaxies (78\%) were the actual target source of
the HRI observation. Thus nearby galaxies, known to be X-ray bright on
the basis of pre-\ro observations, will be well represented in the
XHFS sample, as will established X-ray luminous categories of object
such as Seyfert galaxies.  Table~\ref{tab1} gives details of how the
full HFS sample of galaxies are distributed by morphological type and
nuclear properties and, for comparison, provides the same information
for the XHFS subset. In terms of galaxy morphology, late-type
spirals/irregular galaxies appear to be somewhat favoured. Also, as
expected, a relatively high fraction of the Seyfert galaxies in the
HFS sample were observed by the HRI ({\it i.e.} a $29\%$ fraction
compared to a norm for all types of galaxy of $\sim 17\%$), although
in numerical terms the excess coverage amounts to only about 6
objects. Despite these clear emphases in the XHFS sample, the
different morphological types and nuclear classes are, nevertheless,
reasonable well sampled by the HRI database. We conclude that the XHFS
sample is representative of the range of galaxies that occupy the
nearby Universe (except for giant ellipticals and lenticulars)
although possibly subject to a bias towards high X-ray luminosity
systems.

\begin{table*}
\caption{The distribution with morphological and nuclear type in the full 
HFS sample compared to that in the XHFS subset.}
\centering
\begin{tabular}{lcccccclccc}\hline
\multicolumn{5}{c}{Galaxy Morphology}	& & & \multicolumn{4}{c}{Type of 
Nucleus} \\
Hubble type	& $T$			& \multicolumn{3}{c}{Number of Galaxies}& & & Classification	& \multicolumn{3}{c}{Number of Galaxies} \\
		&			& HFS	& XHFS	& Coverage		& & & 		& HFS	& XHFS	& Coverage	\\\hline
E		& -6 $\rightarrow$ -4	& 57	& 9	& 0.16			& & & Seyferts	& 52	& 15	& 0.29\\
SO		& -3 $\rightarrow$ -1	& 88	& 10	& 0.11			& & & LINERs	& 159	& 26	& 0.16\\
SO/a - Sab	& 0 $\rightarrow$ 2	& 77	& 15	& 0.19			& & & ~~(pure	& 94	& 15	& 0.16)\\
Sb - Sbc	& 3 $\rightarrow$ 4	& 103	& 18	& 0.17			& & & ~~(transition&65	& 11	& 0.17)\\
Sc - Scd	& 5 $\rightarrow$ 6	& 109	& 16	& 0.15			& & & \hii	& 206	& 32	& 0.16\\
Sd and later	& 7 $\rightarrow$ 99	& 52	& 15	& 0.29			& & & NoEL 	& 69	& 10	& 0.14\\
Total		&			& 486	& 83	& 0.17			& & & Total	& 486	& 83	& 0.17\\\hline
\end{tabular}
\label{tab1}
\end{table*}

The individual galaxies contained within the XHFS sample are listed in
Table~\ref{tab2}, together with brief details relating to the \ro HRI
observations and the galaxy properties.

\begin{table*}
\centering
\caption{Details of the 83 galaxies in the XHFS sample.  Repeat
observations of a galaxy are ordered by epoch and labelled
alphabetically.  Values of the distance, $d$, Hubble type, $T$, and nuclear
class are taken from Ho, Filippenko \& Sargent (1997a). The foreground 
(Galactic) column density, \nh, in the direction of each galaxy is based on an
interpolation of the HI measurements of Stark et al. (1992).}
\begin{tabular}{lccccccccc}\hline
  Galaxy    & Obs & \ro data   &  Exposure & $d$ &  \nh  & Hubble & Nuclear& Total & Nuclear \\
 & N$\underline{o}$ &             &  (s) & (Mpc)     & ($10^{20}\rm~cm^{-2}$) & type, $T$  & class   & XRS & XRS?\\\hline
  IC 10   & A & rh600902n00 &  32678 &  1.3 & 48.1&  10.0 & H      &  1   &     \\
          & B & rh600902a01 &  38564 &	    &      &       &	    &      &	 \\
  NGC 147 &   & rh400744n00 &  14745 &  0.7 & 10.5&  -5.0 & 	    &   -  &	 \\
  NGC 185 &   & rh400743n00 &  21070 &  0.7 & 11.0&  -5.0 & S2     &  1   &     \\
  NGC 205 &   & rh600816n00 &  28145 &  0.7 &  6.7&  -5.0 & 	    &   2  &	 \\
  NGC 221 &   & rh600600n00 &  12658 &  0.7 &  6.5&  -6.0 & 	    &   2  &   y \\
  NGC 404 &   & rh703894n00 &  23874 &  2.4 &  5.1&  -3.0 & L2     &  1   &  y	 \\
  NGC 520 & A & rh600628n00 &  13454 & 27.8 &  3.3&  99.0 & H	    &   -  &     \\
          & B & rh600628a01 &  18538 &	    &      &       &	    &      &	 \\
  NGC 891 &   & rh600690n00 &  98118 &  9.6 &  6.8&  3.0  & H	    &   5  &	 \\
  NGC 1058&   & rh500450n00 &  60796 &  9.1 &  5.6&  5.0  & S2     &  -   &     \\
  IC 342  &   & rh600022n00 &  19147 &  3.0 & 30.3&  6.0  & H	    &   7  &   y \\
  NGC 1560&   & rh702727n00 &  17501 &  3.0 & 11.4&  7.0  & H	    &   -  &	 \\
  NGC 1569&   & rh600157n00 &  11350 &  1.6 & 24.2&  10.0 & H	    &   1  &   y \\
  NGC 1961&   & rh600499n00 &  82704 & 53.1 &  8.5&  5.0  & L2     &  1   &  y	 \\
  NGC 2276& A & rh600498n00 &  52235 & 36.8 &  6.0&  5.0  & H	    &   1  &	 \\
          & B & rh600498a01 &  21731 &	    &      &       &	    &      &	 \\
  NGC 2366&   & rh702732n00 &  31823 &  2.9 &  3.9&  10.0 & 	    &   -  &	 \\
  NGC 2403&   & rh600767n00 &  26544 &  4.2 &  4.1&  6.0  & H	    &   4  &	 \\
  NGC 2775& A & rh500385n00 &  10892 & 17.0 &  4.3&  2.0  & 	    &   -  &	 \\
          & B & rh600826a01 &  48468 &	    &      &       &	    &      &     \\
  NGC 2782&   & rh700462n00 &  21715 & 37.3 &  1.8&  1.0  & H	    &   2  &   y \\
  NGC 2903&   & rh600602n00 &  13619 &  6.3 &  3.4&  4.0  & H	    &   3  &   y \\
  NGC 2976& A & rh600471n00 &  18801 &  2.1 &  4.5&  5.0  & H	    &   1  &	 \\
          & B & rh600759n00 &  26524 &	    &      &       &	    &      &	 \\
          & C & rh600759a01 &  23443 &	    &      &       &	    &      &	 \\
  NGC 3031& A & rh600247n00 &  26558 &  1.4 &  4.3&  2.0  & S1.5   &  20  &  y	 \\
          & B & rh600247a01 &  21263 &	    &      &       &	    &      &	 \\
          & C & rh600739n00 &  20080 &	    &      &       &	    &      &	 \\
          & D & rh600740n00 &  19163 &	    &      &       &	    &      &	 \\
          & E & rh600881n00 &  14934 &	    &      &       &	    &      &	 \\
          & F & rh600882n00 &  18508 &	    &      &       &	    &      &     \\
          & G & rh601001n00 &  19420 &	    &      &       &	    &      &	 \\
  NGC 3079& A & rh700100n00 &  21005 & 20.4 &  0.9&  7.0  & S2     &  1   &  y	 \\
          & B & rh600411n00 &  20676 &	    &      &       &	    &      &	 \\
          & C & rh700889n00 &  18677 &	    &      &       &	    &      &	 \\
          & D & rh701294n00 &  25330 &	    &      &       &	    &      &	 \\
          & E & rh702425n00 &  25077 &	    &      &       &	    &      &	 \\
  NGC 3073&   & rh600411n00 &  20676 & 19.3 &  0.9&  -2.5 & H	    &   -  &	 \\
  NGC 3147& A & rh600614n00 &  23438 & 40.9 &  3.2&  4.0  & S2     &  1   &  y	 \\
          & B & rh600721n00 &  22582 &	    &      &       &	    &      &	 \\
          & C & rh600721a01 &  26308 &	    &      &       &	    &      &     \\
  NGC 3185&   & rh800844n00 &  29237 & 21.3 &  2.2&  1.0  & S2:    &  -   &	 \\
  NGC 3190&   & rh800844n00 &  29237 & 22.4 &  2.1&  1.0  & L2     &  1   &  y	 \\
  NGC 3193&   & rh800844n00 &  29237 & 23.2 &  2.1&  -5.0 & L2:    &  -   &	 \\
  NGC 3226&   & rh701299n00 &  30264 & 23.4 &  2.3&  -5.0 & L1.9   &  1   &  y	 \\
  NGC 3227&   & rh701299n00 &  30264 & 20.6 &  2.3&  1.0  & S1.5   &  1   &  y	 \\
  NGC 3294&   & rh500389n00 &  11288 & 26.7 &  1.6&  5.0  & H	    &   -  &	 \\
  NGC 3310&   & rh600685a01 &  41842 & 18.7 &  1.1&  4.0  & H	    &   2  &	 \\
  NGC 3377&   & rh600830n00 &  34272 &  8.1 &  2.9&  -5.0 & 	    &   2  &   y \\
  NGC 3379&   & rh600829n00 &  24560 &  8.1 &  2.9&  -5.0 & L2/T2::&  1   &  y	 \\
  NGC 3384&   & rh600829n00 &  24560 &  8.1 &  2.9&  -3.0 & 	    &   -  &	 \\
  NGC 3389&   & rh600829n00 &  24560 & 22.5 &  2.8&  5.0  & H	    &   -  &	 \\
  NGC 3395&   & rh600771n00 &  24925 & 27.4 &  1.8&  6.0  & H	    &   1  &     \\
  NGC 3628&   & rh700009n00 &  13574 &  7.7 &  2.0&  3.0  & T2     &  2   &	 \\
  NGC 3998&   & rh600430n00 &  13622 & 21.6 &  1.2&  -2.0 & L1.9   &  2   &  y	 \\
  NGC 4051&   & rh701298n00 &  10579 & 17.0 &  1.3&  4.0  & S1.2   &  1   &  y	 \\
  NGC 4088&   & rh500391n00 &  10017 & 17.0 &  2.0&  4.0  & H	    &   1  &   y \\
  NGC 4150&   & rh600762a01 &  10457 &  9.7 &  1.6&  -2.0 & T2     &  1   &  y	 \\
\end{tabular}
\end{table*}

\begin{table*}
\centering
\begin{tabular}{lccccccccc}\hline
  Galaxy    & Obs & \ro data   &  Exposure & $d$ &  \nh  & Hubble & Nuclear& Total & Nuclear \\
 & N$\underline{o}$ &             &  (s) & (Mpc)     & ($10^{20} \rm~cm^{-2}$) & type  & class   & XRS & XRS?\\\hline
  NGC 4151& A & rh701707n00 & 108216 & 20.3 &  2.1&  2.0  & S1.5  &  3  &  y \\
          & B & rh701989n00 & 103435 &	    &      &       &	  &     &    \\
  NGC 4203&   & rh600221n00 &  25447 &  9.7 &  1.3&  -3.0 & L1.9  &  1  &  y \\
  NGC 4214&   & rh600741n00 &  42584 &  3.5 &  1.7&  10.0 & H	  &  1  &    \\
  NGC 4235&   & rh600412a01 &  19694 & 35.1 &  1.5&  1.0  & S1.2   &  1   &  y	 \\
  NGC 4258& A & rh701008n00 &  27556 &  6.8 &  1.4&  4.0  & S1.9   &  8  &  y  \\
          & B & rh600599a01 &  25562 &	    &      &       &	    &      &	 \\
  NGC 4291& A & rh600441n00 &  12405 & 29.4 &  3.2&  -5.0 & 	    &   1  &   y \\
          & B & rh600834n00 &  35363 &	    &      &       &	    &      &	 \\
  NGC 4321&   & rh600731n00 &  42797 & 16.8 &  2.4&  4.0  & T2     &  4   &  y	 \\
  NGC 4388&   & rh700192n00 &  11274 & 16.8 &  2.7&  3.0  & S1.9   &  1   &  y	 \\
  NGC 4395&   & rh702725n00 &  11353 &  3.6 &  1.3&  9.0  & S1.8   &  1   &	 \\
  NGC 4435&   & rh600608n00 &  21651 & 16.8 &  2.7&  -2.0 & T2/H   &  1   &  y	 \\
  NGC 4438&   & rh600608n00 &  21651 & 16.8 &  2.7&  0.0  & L1.9   &  1   &  y	 \\
  NGC 4449& A & rh600743n00 &  15373 &  3.0 &  1.3&  10.0 & H	   & 7  &  y\\
          & B & rh600865n00 &  25395 &	    &      &       &	    &      &	   \\
          & C & rh600865a01 &  18343 &	    &      &       &	    &      &	   \\
  NGC 4470&   & rh600216a01 &  27367 & 31.4 &  1.6&  1.0  & H	    &   -  &	   \\
  NGC 4485& A & rh600855n00 &  27048 &  9.3 &  1.8&  10.0 & H	    &   1  &   y   \\
          & B & rh600855a01 &  24829 &	    &      &       &	    &      &	   \\
  NGC 4490& A & rh600855n00 &  27048 &  7.8 &  1.8&  7.0  & H	    &   4  &   y   \\
          & B & rh600855a01 &  24829 &	    &      &       &	    &      &	   \\
  NGC 4527&   & rh500390n00 &  10355 & 13.5 &  1.9&  4.0  & T2     &  -   &	   \\
  NGC 4559&   & rh600861n00 &  53812 &  9.7 &  1.5&  6.0  & H	    &   5  &   y   \\
  NGC 4569&   & rh600603a01 &  21858 & 16.8 &  2.5&  2.0  & T2     &  1   &  y	   \\
  NGC 4631& A & rh600193a01 &  14287 &  6.9 &  1.3&  7.0  & H	    &   2  &	   \\
          & B & rh600193a02 &  12041 &	    &      &       &	    &      &	   \\
  NGC 4638& A & rh600480a01 &  11915 & 16.8 &  2.4&  -3.0 & 	    &   -  &	   \\
          & B & rh600620a01 &  19923 &	    &      &       &	    &      &	   \\
  NGC 4647& A & rh600480a01 &  11915 & 16.8 &  2.4&  5.0  & H	    &   -  &	   \\
          & B & rh600620a01 &  19923 &	    &      &       &	    &      &	   \\
  NGC 4651&   & rh800719n00 &  25396 & 16.8 &  2.0&  5.0  & L2     &  1   &  y	   \\
  NGC 4656&   & rh600605n00 &  27668 &  7.2 &  1.3&  9.0  & H	    &   1  &     \\
  NGC 4736& A & rh600678n00 & 112910 &  4.3 &  1.4&  2.0  & L2     &  5   &  y	 \\
          & B & rh600769n00 &  27292 &	    &      &       &	    &      &	 \\
  NGC 4772&   & rh500397n00 &  12120 & 16.3 &  1.8&  1.0  & L1.9   &  -   &	 \\
  NGC 4826&   & rh600715n00 &  10150 &  4.1 &  2.6&  2.0  & T2     &  1   &  y	 \\
  NGC 5005&   & rh701711n00 &  26742 & 21.3 &  1.1&  4.0  & L1.9   &  1   &  y	 \\
  NGC 5055&   & rh600742n00 &  12343 &  7.2 &  1.4&  4.0  & T2     &  9   &  y	 \\
  NGC 5194&   & rh600601n00 &  36323 &  7.7 &  1.5&  4.0  & S2     &  9   &  y	 \\
  NGC 5195&   & rh600601n00 &  36323 &  9.3 &  1.5&  90.0 & L2:    &  1   &	 \\
  NGC 5204& A & rh702723n00 &  14852 &  4.8 &  1.5&  9.0  & H	    &   1  &   y \\
          & B & rh702723a01 &  13675 &	    &      &       &	    &      &	 \\
  NGC 5273&   & rh701006n00 &  10767 & 21.3 &  0.9&  -2.0 & S1.5   &  1   &  y	 \\
  NGC 5354&   & rh800743n00 &  22761 & 32.8 &  1.0&  -2.0 & T2/L2: &  -   &	 \\
  NGC 5457& A & rh600092n00 &  18588 &  5.4 &  1.2&  6.0  & H	    &   27 &	 \\
          & B & rh600383n00 &  32624 &	    &      &       &	    &      &	 \\
          & C & rh600820n00 & 108850 &	    &      &       &	    &      &	 \\
          & D & rh600820a01 &  68872 &	    &      &       &	    &      &	 \\
  NGC 5775&   & rh600964n00 &  35305 & 26.7 &  3.3&  5.0  & H	    &   -  &	 \\
  NGC 5850& A & rh600478n00 &  22796 & 28.5 &  4.2&  3.0  & L2     &  -   &	 \\
          & B & rh600478a01 &  16763 &	    &      &       &	    &      &	 \\
  NGC 5905&   & rh703855n00 &  76218 & 44.4 &  1.4&  3.0  & H	    &   1  &   y \\
  NGC 6217& A & rh141910n00 &  15991 & 23.9 &  4.1&  4.0  & H	    &   1  &   y \\
          & B & rh141913n00 &  14669 &	    &      &       &	    &      &	 \\
          & C & rh141921n00 &  13572 &	    &      &       &	    &      &	 \\
          & D & rh141926n00 &  11804 &	    &      &       &	    &      &	 \\
          & E & rh141928n00 &  11630 &	    &      &       &	    &      &	 \\
  NGC 6503&   & rh600618n00 &  14787 &  6.1 &  4.0&  6.0  & T2/S2: &  1   & 	 \\
  NGC 6654& A & rh600124a00 &  36803 & 29.5 &  5.4&  0.0  & 	    &   -  &	 \\
          & B & rh600124a01 &  18270 &	    &      &       &	    &      &	 \\
  NGC 6946& A & rh600501n00 &  60305 &  5.5 & 20.2&  6.0  & H	    &   13 &   y \\
          & B & rh600718n00 &  21723 &	    &      &       &	    &      &	 \\
  NGC 7331&   & rh702074n00 &  30481 & 14.3 &  7.7&  3.0  & T2     &  1   &  y	 \\\hline
\end{tabular}
\label{tab2}
\end{table*}

\begin{figure*}
\centering
\includegraphics[width=8cm]{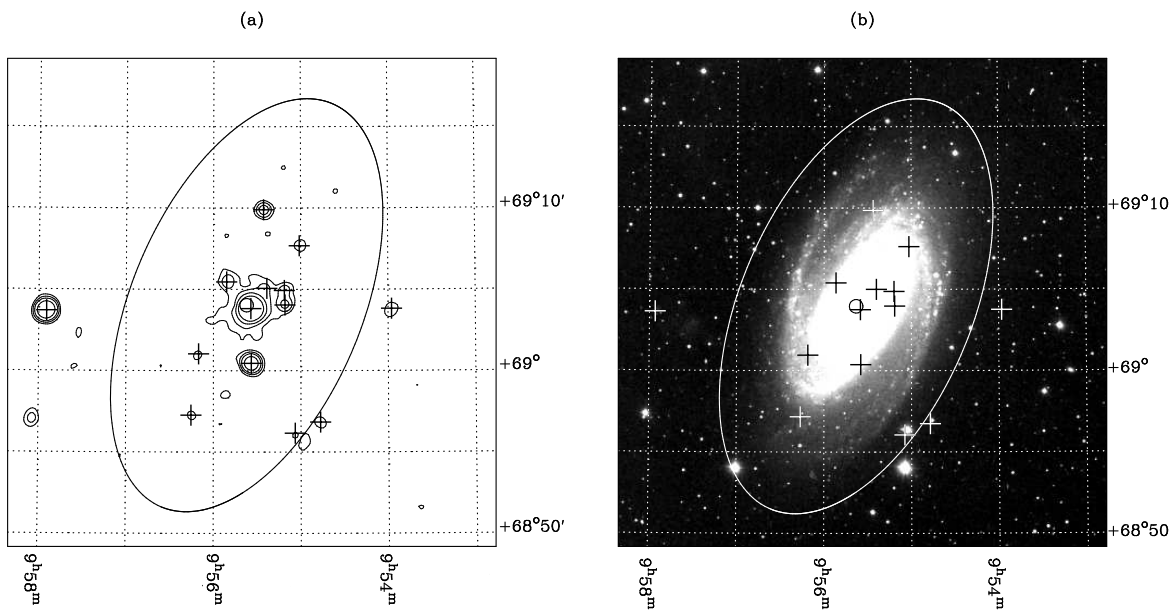}
\caption{(a) A contour map of the image derived from the first \ro HRI
observation of NGC 3031 (M81). The point X-ray sources detected by the
PSS algorithm are marked with crosses. (b) The position of the point
X-ray sources superimposed on the DSS optical image. In both panels
the position of the optical nucleus is marked by the circle (25$''$
radius) and the ellipse represents to the \d25 blue isophote.}
\label{fig0}
\end{figure*}

The point source searching was conducted using the {\small PSS}
algorithm, which is part of the {\small STARLINK ASTERIX} software package
(Allen 1995).  A circular area of 15$'$ radius about the field centre
was searched in each case and all discrete X-ray sources with a
detection significance of $\ge 5~ \sigma$ above a smoothed background
level were recorded.  The source positions were then plotted on the
X-ray image and a visual ``quality'' check of each detected source was
carried out. Sources were flagged as ``on-galaxy'' or ``off-galaxy''
dependent upon whether they resided inside or outside of the optical
extent of the galaxy. The optical extent was defined in terms of the
\d25 ellipse (\ie the elliptical contour best corresponding to the 25
magnitude/arcsec$^{2}$ blue isophote) using the major axis diameter
and nuclear position from Ho, Filippenko \& Sargent (1997a), and the
position angle and axial ratio from the Third Reference Catalogue of
Bright Galaxies (RC3; de Vaucouleurs et al. 1991).  In this manner a
total of 327 discrete X-ray sources were detected coincident with the
galaxies in our sample.  Figure~\ref{fig0} illustrates this process in
the case of NGC 3031 (M81).

Since a considerable proportion (24/83) of our galaxies were observed
on at least two separate occasions, many of these detections
represented repeat observations of the same X-ray sources.  We
therefore compressed the source list to a ``definitive'' list of
detections by considering all sources positionally coincident to
within 15$''$ in separate observations of the same galaxy, to be the
same source (see \S 3.1)\footnote{We note that in correlating the
multiple detections we only needed to apply an attitude correction to
one field, namely NGC 4258 A, which required corrections of
$\Delta$RA$= +1.35^s$, $\Delta$Dec$= +16.5''$ (calibrated by a bright
off-axis source).}.  This led to a final set of 187 discrete X-ray
sources, the catalogue of which is presented in Appendix A.

Visual inspection suggests the presence of diffuse X-ray flux in a
small fraction of the HRI images considered above. Most notably there
are five cases where a bright point-like X-ray source, positionally
coincident with the optical nucleus of the galaxy, appears to be
embedded in an extended, low surface brightness component. NGC 3031 in
Figure 1(a) provides one such example, the others being NGC 4151, NGC
4258, NGC 4291 and NGC 4736.  We have not specifically flagged
extended emission in our analysis and, in fact, ignore its
contribution to the X-ray luminosity of the XHFS galaxies, focussing
our attention solely on individual discrete (point-like) X-ray sources
associated with nearby galaxies. (In all cases any correction for the
the extended luminosity {\it detected by the HRI} would be much less
than a factor 2).


All fluxes (and hence luminosities) quoted in this paper are for the
0.1--2.4 keV \ro HRI band, corrected for the \los absorption of our
Galaxy.  The conversion from measured HRI count rate to unabsorbed
flux was calculated for a specific spectral form, namely a power-law
continuum with photon index $\Gamma = 2$.  The absorption correction
is, in most cases, small (less than a factor 2 upwards) since the
galaxies in our sample are generally in regions of low Galactic column
density (\nh~$< 5 \times 10^{20} \atpcm$). The count rate to flux
conversion factor is sensitive to the choice of spectral model, for
example a change to a thermal bremsstrahlung form with $kT = 1$ keV
leads to a $20\%$ reduction in the assigned fluxes.

\section{Preliminary Analysis of the XHFS Sample}

\subsection{Nuclear versus non-nuclear source designation}

As a first step in the analysis we have attempted to separate out
those X-ray sources possibly associated with the nuclei of the XHFS
galaxies from the other constituent X-ray source populations.  The
reason for this is that an active galactic nucleus (AGN) is known or
suspected to be present in quite a number of the galaxies in our
sample and, more often than not, is the dominant source of X-ray
emission in the nuclear region and sometimes in the whole galaxy
(examples include NGC 4736, Roberts, Warwick \& Ohashi 1999; NGC 3031,
Ishisaki et al. 1996; NGC 3147, Ptak et al. 1997).  Since, as far as
we know, galaxies contain at most one AGN located close to the mass
centroid of the galaxy, then there is clearly some logic in
differentiating between ``nuclear'' and ``non-nuclear'' sources.

In the present study the identification of a particular X-ray source
with the nucleus of a galaxy is based solely on a positional
coincidence criterion.  In \ro HRI observations there is the potential
for systematic pointing errors of up to $\sim 10''$ (see Harris et
al. 1998) in addition to statistical errors of up to $\sim 5''$
(dependent on the source strength). For example, in the case of NGC
3079, we observe the position of one particular source (designated NGC
3079 X-1 in Appendix A) to shift by up to 15$''$ in 5 separate
observations.  There is also some uncertainty in the position of the
optical nucleus; Ho, Filippenko \& Sargent (1995) only quote their
nuclear positions to one second of time implying a potential error of
up to $7.5''$.  This would suggest that a reasonable criterion for
designating an X-ray source as ``nuclear'' would be if it lies within
$\sim 20''$ of the optical nucleus.  In fact the observations reveal a
total of 40 X-ray sources with an offset from the optical nucleus of
less than $20''$, with a further 5 sources lying in the 20 - 25$''$
range.  However, 4 out of these 5 sources have an X-ray luminosity in
excess of the median value for all the galactic nuclei and,
considering this along with the detection of some probable AGN out to
$\sim 20''$ from the optical nuclear position, we therefore set a very
{\it conservative} limit of $25''$ as a threshold radius (thereby
avoiding any possible contamination of the non-nuclear sample by AGN).
Hereafter we refer to any sources seen within this bound as
``nuclear'' X-ray sources and label those outside this region as
``non-nuclear'' X-ray sources.



\begin{figure*}
\centering
\includegraphics[width=5cm]{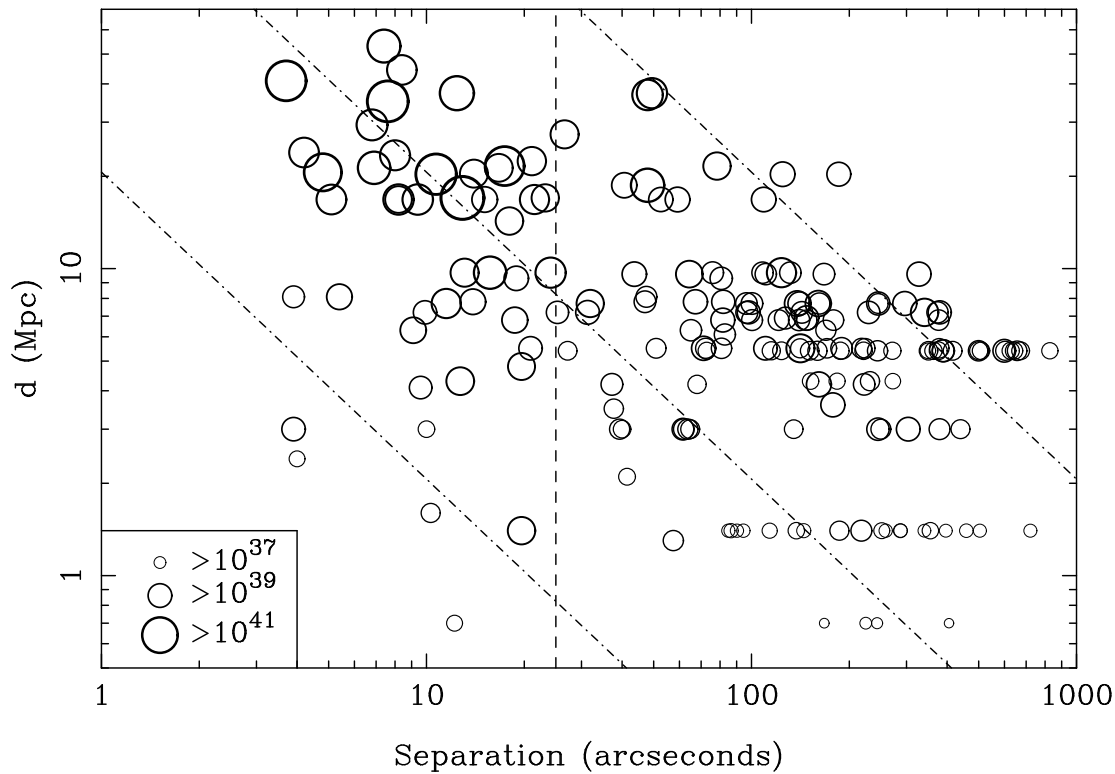}
\caption{The observed separation of the 187 X-ray sources from the
position of the optical nucleus of the host galaxy.  The vertical
(dashed) line drawn at a separation of 25$''$ divides the sources
classified as ``nuclear'' from those considered to be ``non-nuclear''.
The three diagonal (dot-dash) lines correspond to projected linear
scales of 100 pc, 1 kpc and 10 kpc.  The luminosity of each source is
represented by the size of its circle as illustrated in the inset.}
\label{fig1}
\end{figure*}

The division of the XHFS sample into nuclear and non-nuclear X-ray
sources is illustrated in Figure~\ref{fig1}. One point that is
apparent immediately from this figure is that relatively high
luminosity (\lx~$\ga 10^{39} \rm~erg~s^{-1}$) sources are not confined
solely to the nuclear region, in fact such sources are seen at
projected distances from the optical nucleus ranging from less than a
few hundred pc (\ie positional coincident with the nucleus in terms of
the measurement errors) out to beyond 10 kpc. In a recent paper
exploiting \ro HRI observations, Colbert \& Mushotzky (1999), have
suggested the possibility that some nearby spiral galaxies contain
X-ray sources powered by accretion onto black holes with masses of
typically $10^{2} - 10^{4} \Msun $.  These authors invoke the
existence of a black-hole population in this intermediate mass range
(compared to the Galactic black-hole binaries with masses typically $
\sim 10 \Msun$ and luminous AGN with $ M \sim 10^{6} - 10^{9} \Msun $)
to explain why X-ray sources with luminosities of \lx $\sim 10^{40}
\rm~erg~s^{-1}$ are occasionally observed in spiral galaxies with a
significant positional offset from the optical nucleus.  In fact
Colbert \& Mushotzky quote a mean offset of $\sim 390$ pc between the
X-ray source position and the optical photometric center for the 21
galaxies with X-ray detections in their sample\footnote{A total of 11
out of the 39 galaxies studied by Colbert \& Mushotzky (1999) overlap
with our XHFS sample}.  In Figure~\ref{fig1} there is no real evidence
for the more luminous sources to be grouped preferentially within a
projected separation of 1 kpc but such an effect, if present, might
well be smeared out by the positional errors. Colbert \& Mushotzky
(1999) concentrate on a set of galaxies with recessional velocity less
than $1000\rm~km~s^{-1}$ (\ie a {\it very nearby} sample) and hence
benefit from the better effective linear resolution.

We find that 45 galaxies out of 83 in the XHFS sample harbour a
nuclear X-ray source.  The corresponding statistic for the non-nuclear
X-ray sources is that 34 of the XHFS galaxies contain at least one
such source.  There were no X-ray detections in 21 of the XHFS
galaxies.

\begin{figure*}
\centering
\includegraphics[width=14cm]{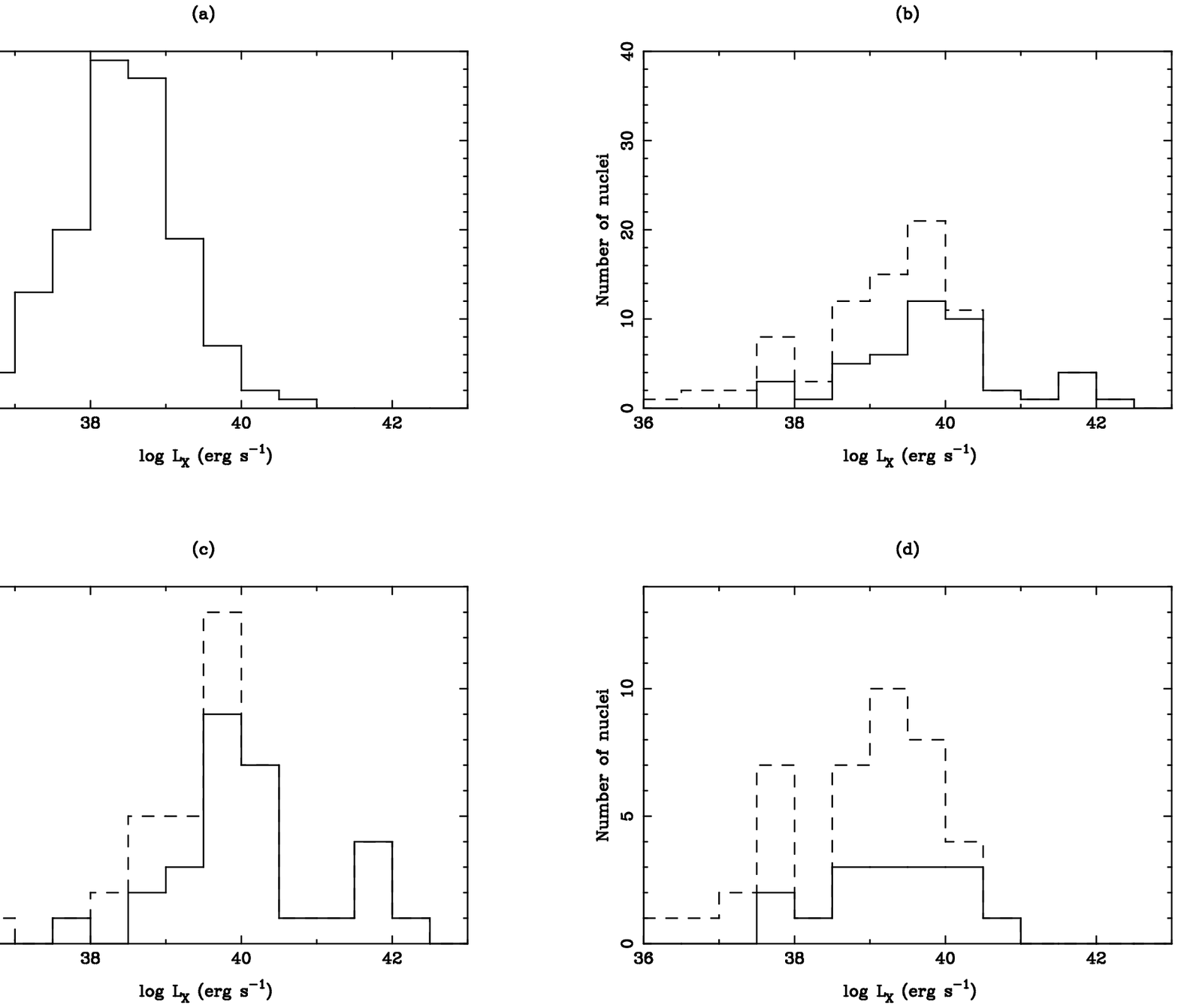}
\caption{The distribution of X-ray luminosity exhibited by sources in
the XHFS sample. (a) The non-nuclear sources. (b) The nuclear
sources. The dashed histogram shows the effect of including the 95\%
upper limits for the galaxies with no detected nuclear X-ray
source. (c) Same as (b) but restricted to galaxies with Seyfert or
LINER nuclei. (d) Same as (b) but restricted to galaxies with \hii
nuclei or no nuclear emission lines.}
\label{fig2}
\end{figure*}

In Figure~\ref{fig2} (panels a \& b) we present histograms of the
observed luminosity distribution for the 187 sources in the XHFS
sample separated into the non-nuclear and nuclear categories.
Figure~\ref{fig2}(b) also includes 95\% upper limits on the X-ray
luminosity of the nuclear emission for the 38 galaxies with no
detected nuclear X-ray source (these upper limits are tabulated in
Appendix B). Although the luminosity distributions for the two populations 
overlap, on average the {\it detected} nuclear X-ray 
sources are substantially more X-ray luminous than the non-nuclear sources
(the median X-ray luminosity of the former being $5.4 \times 10^{39} \ergsec$
as opposed to $2.8 \times 10^{38} \ergsec$ for the latter).

It should be emphasized that the X-ray sources designated as nuclear
are not necessarily all low-luminosity AGN. Indeed, since the range of
X-ray luminosity exhibited by the non-nuclear sources shows a
significant overlap with the nuclear sample, some contamination of the
latter category by non-AGN looks inevitable.  The problem of
distinguishing very low-luminosity AGN from individual black hole
X-ray binaries, groupings of Eddington-limited low-mass XRBs and
recent X-ray luminous supernovae is, of course, well recognised (\eg
Dahlem, Heckman \& Fabbiano 1995; Loewenstein et al. 1998; Lira et
al. 1999). However the above argument does not preclude the
possibility that many of the nuclear X-ray sources contained in the
XHFS sample are putative low-luminosity AGN, particularly if the
underlying population of super-massive black holes have a very wide
range of masses and/or accretion rates (Colbert \& Mushotzky 1999).

\subsection{Contamination of the XHFS sample by background 
QSOs and/or foreground stars}

A potential source of uncertainty is the detection of either
foreground sources or background sources which lie, by chance, within
the \d25 ellipse of one of the XHFS galaxies. Such contamination is
likely to be much more of a problem for the non-nuclear category
since, in this case, the total sky area surveyed amounts to almost
exactly one square degree (with the area within the nuclear error
circles accounting for only $\sim 1\%$ of this figure).  At the flux
limits reached in the present survey (see below), most of the
contamination is due to late-type stars in our own Galaxy and distant
QSOs (the latter representing the dominant X-ray source population in
medium-deep \ro images).

\begin{figure*}
\centering
\includegraphics[width=14cm]{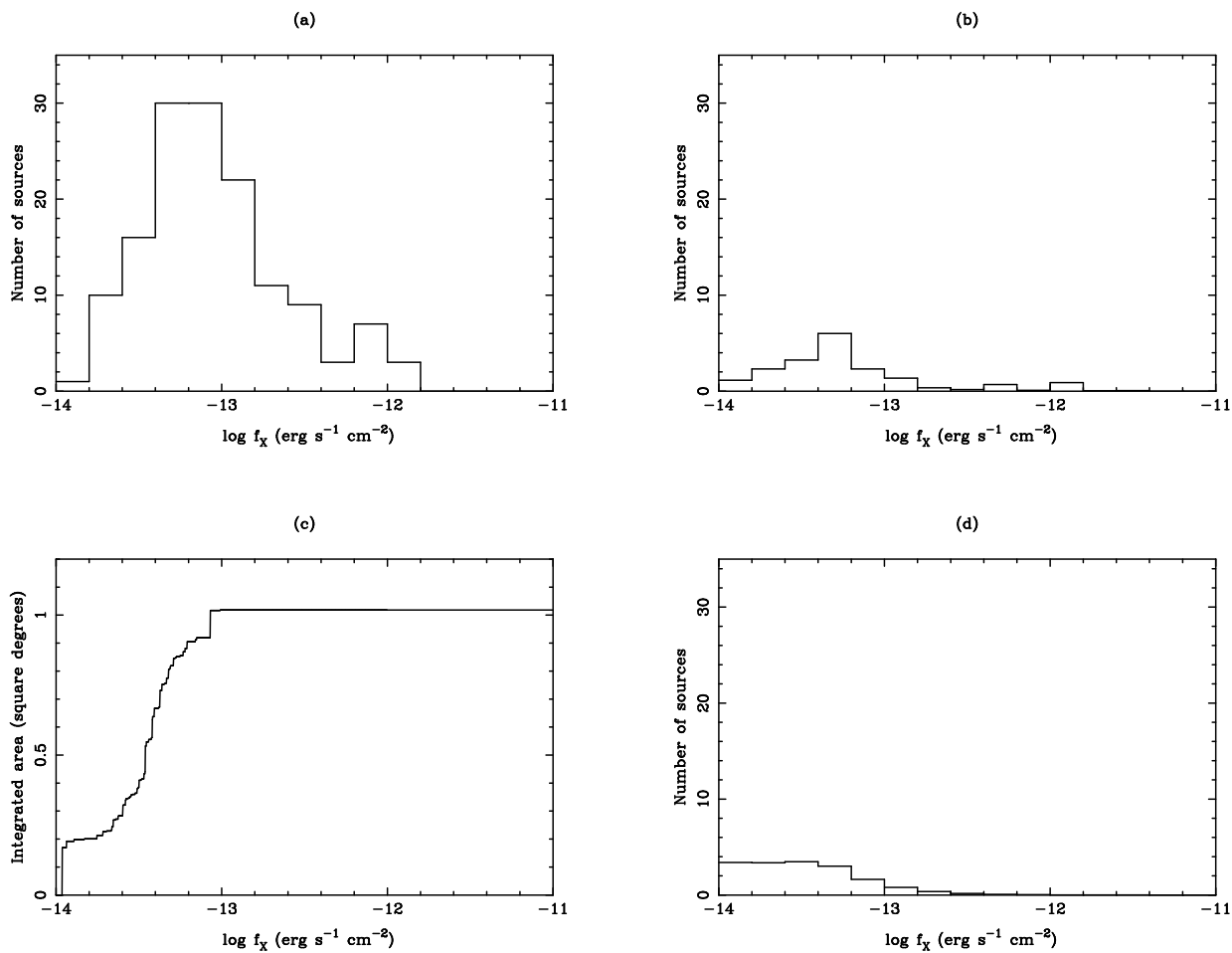}
\caption{(a) The observed flux distribution of the non-nuclear X-ray
sources.  (b) The predicted flux distribution for contaminating
sources based on the off-galaxy detections.  (c) The sky coverage
(within the galaxy \d25 contours) versus the limiting flux.  (d) The
predicted flux distribution of contaminating sources derived from the
differential X-ray source counts of Hasinger et al. (1993).}
\label{fig3}
\end{figure*}

In Figure~\ref{fig3}(a) we show the distribution of non-nuclear source
detections as a function of the observed X-ray flux. We have carried
out a quantitative investigation of the contamination of the
non-nuclear population expected at different flux levels using two
independent approaches.  Firstly we estimated the level of
contamination in each HRI field using the detections outside of the
\d25 ellipse of the target galaxy but within the nominal HRI field of
view as a control set. After applying an appropriate field by field
scaling of the number of off-galaxy detections\footnote{The weighting
factor was simply the ratio of the on-galaxy to off-galaxy area for
the field; here and elsewhere in this paper we neglect vignetting and
other effects which reduce the point-source sensitivity at the edge of
the HRI field of view compared to that at the centre by about
$10\%$.}, we obtain an estimate of the number of false associations
expected as a function of flux as illustrated in Figure~\ref{fig3}(b).
Comparison of the observed distribution with this estimate
demonstrates that the contamination of the non-nuclear XHFS sample is
at a level of no more than about $13\%$.  As an additional check we
also performed a calculation based on published source counts for the
soft X-ray band.  First the X-ray source detection limit for each field 
was estimated and the integrated sky area encompassed in the
XHFS (within the \d25 ellipses of the XHFS galaxies) determined as a
function of the limiting X-ray flux - see Figure~\ref{fig3}(c). 
The differential form of the {\it log N - log S\/} relation derived by
Hasinger et al. (1993), which includes contributions from both QSOs
and Galactic stars, was then folded through this coverage curve to give the
predicted number of sources shown in Figure~\ref{fig3}(d).  The
contamination level determined from the second method is $11\%$, very
comparable, within the statistics, to that obtained from the first.

There are two caveats relating to this comparison of observed and
predicted source numbers. The effect of absorption by cool gas and
dust contained within the galaxy disks will help suppress the
detection rate of background QSOs, implying that both of the estimates
above actually give upper limit predictions. A second possibility is
that the measured off-galaxy detection rate may be biased upwards by
sources actually associated with the target galaxy but lying outside
its \d25 contour. This latter influence could give rise to the slight
differences between the two predicted distributions in
Figure~\ref{fig3}, but clearly the evidence for any such effects is
very marginal.  In summary we can be confident that the great majority
(in excess of 85\%) of the non-nuclear X-ray sources in our sample are
associated with HFS galaxies. We also confirm that there is
essentially no contamination of the nuclear source sample, since we
expect a total of $\sim 0.2$ sources coincident by chance compared to
the 45 sources we detect.

As a final point we note that if we transpose the distributions in
Figure~\ref{fig3}(b,d) into luminosity space using the appropriate
galaxy distances (on a field by field basis), then the distribution in
luminosity (of the contamination) is very comparable to that exhibited
by the non-nuclear sources in Figure~\ref{fig2}(a). We conclude that
the high luminosity tail of non-nuclear sources is not due to
contamination by foreground or background sources but represents a
real super-luminous population associated with nearby galaxies.

\section{Properties of the nuclear sample}

The search for ``dwarf'' Seyfert nuclei carried out by Ho, Filippenko
\& Sargent (1995, 1997a,b,c) has resulted in a consistent and complete
classification of the optical emission line properties of each
galactic nucleus in the HFS galaxy sample.  They found that 11\% of
nearby galaxies harbour a Seyfert-like nucleus, 33\% host a Low
Ionisation Nuclear Emission-line Region (LINER) and 42\% have a
nucleus with an emission line spectrum characteristic of \hii
regions. The LINERs may be further split into the subsets of ``pure''
LINERs (19\% of the HFS sample) and ``transition'' LINERs (14\%), the
difference being that the transition objects display a blend of LINER
and \hii spectra. Only 14\% of galaxies appear to have nuclear spectra
devoid of optical emission lines (which are referred to here as
``NoEL'' nuclei).

As noted earlier (see Table~\ref{tab1}) the XHFS sample encompasses
the full range of nuclear spectral classification, thus allowing us to
consider the X-ray properties, specifically the soft X-ray
luminosities, of the different types of nucleus.  Figure~\ref{fig2}
(panels c \& d) compares the X-ray luminosity distribution for those
galaxies with Seyfert or LINER nuclei to those with \hii spectra or no
emission lines.  Table~\ref{tab3} gives further details of the mean
nuclear X-ray luminosity and detection rate for various sub-groupings 
in the XHFS sample.\footnote{We have calculated the mean value of log \lx
and the associated error using the Kaplan-Meier estimator procedure in the
the {\bf ASURV} software package (Rev 1.2), which takes account of both the
actual measurements and upper-limits (see Feigelson \& Nelson 1985).
Prior to applying the estimator a few upper-limits below log \lx $= 37.5$ 
were reset to this minimum value. In some cases the lowest value in the 
distribution was an upper-limit resulting in a (probably modest) upward 
biasing of the mean.}

\begin{table*}
\caption{The mean X-ray luminosities and detection rates of various 
optically-defined categories of nuclei.}  
\centering
\begin{tabular}{lccc}\hline
Sample	& No. of	& Mean log \lx		& Detection rate \\
	& Objects 	& ($\ergsec$)	& (\%) 		\\\hline
All nuclei	 & 83	& $38.93\pm0.15$ 	& 54 \\
 & & & \\
Seyferts \& LINERs & 41	& $39.51\pm0.22$	& 71 \\
HII \& NoEL     & 42	& $38.36\pm0.17$	& 42 \\
 & & & \\
Seyferts	& 15	& $39.83\pm0.43$	& 73 \\
LINERs		& 26	& $39.36\pm0.22$	& 69 \\
~(pure   	& 15	& $39.49\pm0.32$	& 73) \\
~(transition	& 11	& $39.28\pm0.18$	& 64) \\
\hii		& 32	& $38.48\pm0.20$	& 41 \\
NoEL		& 10	& $38.02\pm0.30$	& 30 \\
 & & & \\
Broad Line AGN  & 15	& $40.38\pm0.30$	& 87 \\
Narrow Line AGN & 26	& $39.05\pm0.24$	& 62 \\\hline
\end{tabular}
\label{tab3}
\end{table*}

For the overall sample the average X-ray luminosity is $\sim 
10^{39} \ergsec$.  However, a marked difference is seen between
those nuclei probably hosting a low-luminosity AGN (the
Seyfert and LINER nuclei) and those showing no evidence for
high-ionization emission lines (the \hii and NoEL galaxies).
Essentially, we observe the former to be considerably more luminous
than the latter.  This manifests itself as both a higher X-ray luminosity 
(by a factor ${\sim} 10 $) and a higher detection rate of
nuclear sources. A formal comparison of the two distribution shown in
Figure~\ref{fig2} (panels c \& d) using the two-sample tests in the 
{\bf ASURV} package confirms that the distributions are different at a 
significance level greater than 99.9\%. 

Within the putative AGN grouping, there is little to distinguish
Seyfert from LINER nuclei in terms of either their average luminosity,
detection rate or the luminosity spread.  When the LINERs are divided
into ``pure'' and ``transition'' types, the former are seen to bear a
slightly closer resemblance to the Seyferts, in terms of their X-ray
characteristics, than the latter (although the differences are at a
marginal significance level).  The galaxies with broad emission lines
(the ``type 1'' AGN), discussed by Ho, Filippenko \& Sargent (1997b),
are clearly the most luminous group of objects in the XHFS sample
having an X-ray luminosity at least an order of magnitude larger than
the narrow-emission line (``type 2'') sources and a higher detection
rate.  Nevertheless the average luminosity of the type 2 objects is
still well in excess of that of the \hii and NoEL galaxies (formally
the log \lx distributions are different at a significance level of
98\%). We conclude that X-ray emission associated with an underlying
low-luminosity AGN is seen in almost all of the nearby galactic nuclei
classified as type 1 Seyferts or LINERs and in a reasonable fraction
($\sim 60\%$) of those nuclei classified as type 2 objects. Since
obscuration in the line of sight to the nucleus in type 2 objects will
strongly suppress the 0.1--2.4 keV flux if the column density is in
excess of $ \sim 10^{22} \rm~cm^{-2}$, it may be that in many of the
type 2 nuclei we actually observe scattered X-ray flux (or some other
nuclear related extended component) rather than the direct continuum
emission of the nucleus.

The observed distribution of nuclear luminosity in the galaxies 
categorised as \hii or NoEL objects overlaps
substantially with that measured for non-nuclear sources in the XHFS
sample (Figure~\ref{fig2}(a) \& (d)). Thus it is quite plausible
that in many of the \hii and NoEL galaxies with nuclear detections we
are, in fact, observing a component of the discrete X-ray source
population of the galaxy rather than the presence of an underlying,
optically faint, AGN.

\begin{figure*}
\centering
\includegraphics[width=10cm]{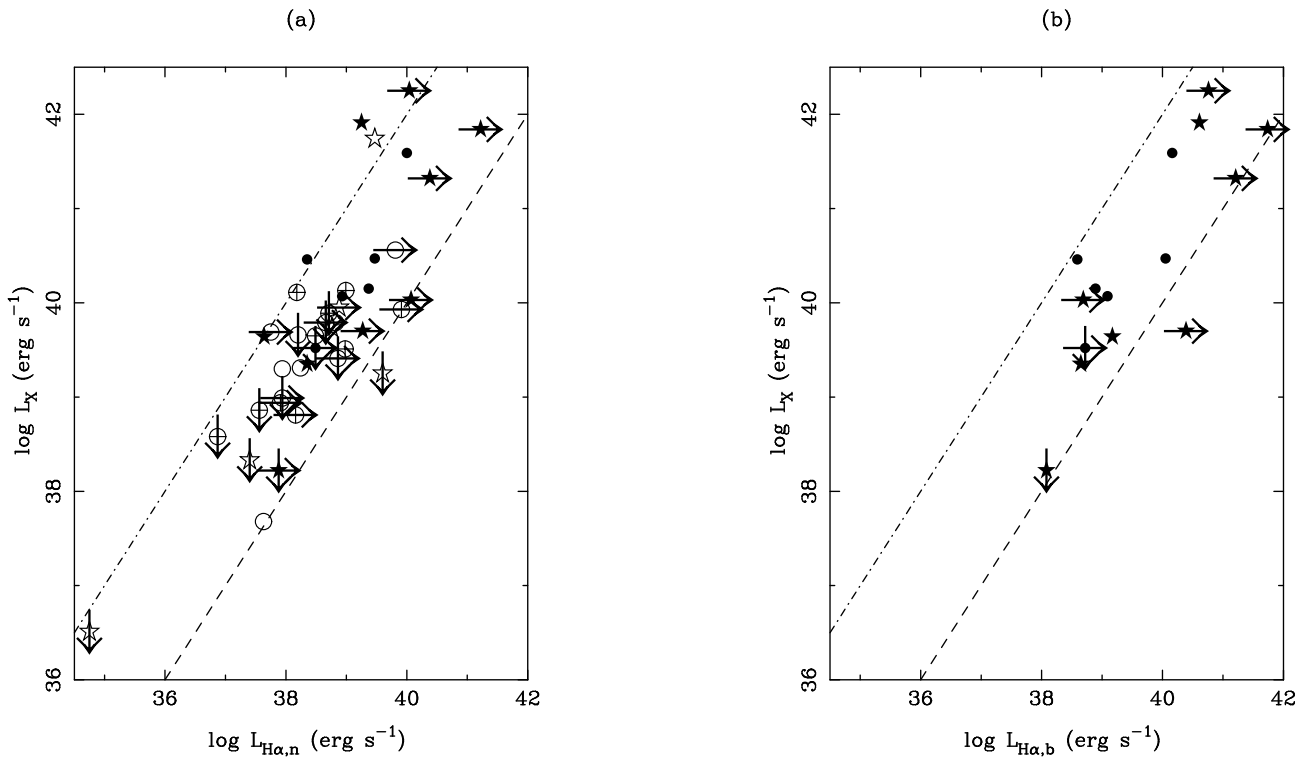}
\caption{Scatter plots of the nuclear X-ray luminosity, \lx~versus the 
nuclear \Ha~luminosity (a) for the narrow component of
\Ha, (b) for the broad component of \Ha.  In
both panels Seyferts are marked by stars and LINERs by circles, with
the broad-line examples of either type of object shown by a filled-in
symbol. The transition LINERs are indicated by a crossed circle.
Limits are indicated by arrows through the symbols.  Ratios of
\lx/L$_{\rm H\alpha}$ of 1 and 100 are shown by the dashed and
dot-dash lines respectively.}
\label{fig4}
\end{figure*}

The properties of the putative AGN in the XHFS sample may be further
investigated by comparing the observed X-ray luminosity with other
indicators of nuclear activity. We examine the relationship between
the X-ray and \Ha~luminosity (both narrow and broad components) for
the Seyferts and LINERs in Figure~\ref{fig4}.  The \Ha~luminosities
(including lower limits on the narrow component, L$_{\rm H\alpha,n}$,
from measurements made in non-photometric conditions) were taken from
Ho, Filippenko \& Sargent (1997a,b); where no measurement of the broad
component, L$_{\rm H\alpha,b}$, is available we use the lower limit on
L$_{\rm H\alpha,n}$ and the fractional contribution of broad \Ha~to
the total \Ha~emission to derive a lower limit on L$_{\rm H\alpha,b}$.

A correlation between the total \Ha~luminosity and soft X-ray
luminosity of high luminosity QSOs and Seyfert I galaxies is well
established, and has recently been shown to extend down to \lx~$\sim
10^{40} \ergsec$ by Koratkar et al. (1995).  They find that type 1 AGN
have a typical \lx/L$_{\rm H\alpha}$ ratio in the range 1 - 100, with
a mean of $\sim 30$.  These results broadly agree with
Figure~\ref{fig4}, where an apparently similar range of ratios is seen
to extend down to $\sim 10^{38} \ergsec$.  Koratkar et al. (1995) show
their correlation is present in flux as well in luminosity space. When
Figure~\ref{fig4} is replotted in this regime, there is little
evidence of a correlation, but this is probably due to the restricted
flux range, the degree of scatter being similar to that seen by
Koratkar et al. The factors which may induce scatter in the X-ray
luminosity at a given \Ha~ output include soft X-ray absorption,
nuclear flux variability and contamination by the non-AGN X-ray source
population of the galaxies (which is clearly a particular problem in
the low-luminosity range).  The present work demonstrates that the
\lx/L$_{\rm H\alpha}$ relationship may extend from \lx~$ > 10^{46}
\ergsec$ down to \lx~$\sim 10^{38} \ergsec$, two magnitudes lower than
previously seen and implies a significant commonality in the physical
processes that drive low- and high-luminosity AGN.

Dynamical studies of the centres of nearby galaxies have revealed the
presence of Massive Dark Objects (MDOs) in the nuclei of number of
galaxies (Kormendy \& Richstone 1995).  The demographics of this
population show that a reasonably tight relationship exists between
the masses of the MDOs and the mass of the galactic bulge hosting them
($\pm 0.5$ decades standard deviation in the ratio; Magorrian et
al. 1998).  Since the mass-to-light relationship of these bulges is
also known then a mass of the MDO may be inferred from the optical
blue bulge luminosity, L$_{\rm bulge}$, of a galaxy.  In the following
analysis we make the assumption that these MDOs are most likely
supermassive black holes resident in the nucleus of the galaxies; this
assumption, and caveats relating to the observations of MDOs, are
discussed by Lawrence (1999, and references therein).

\begin{figure*}
\centering
\includegraphics[width=10cm]{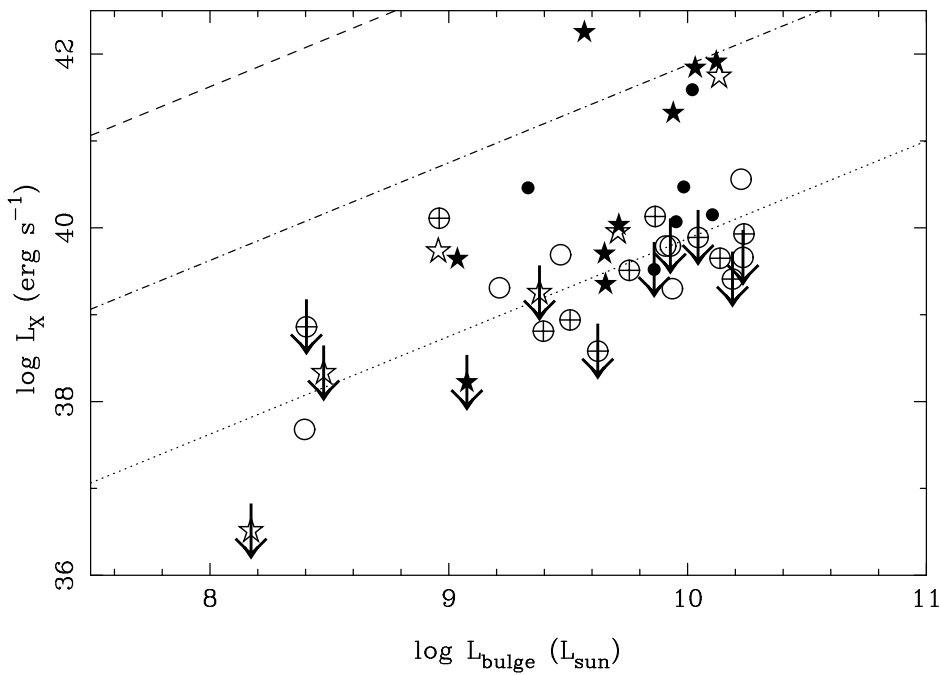}
\caption{The X-ray luminosity of the Seyfert and LINERs in the XHFS sample 
plotted against the optical bulge luminosity of the host galaxy.
The symbols for the various types of AGN are as per Figure~\ref{fig4}.
The diagonal lines represent the $10^{-2}$ (dashed), $10^{-4}$
(dot-dashed) and $10^{-6}$ (dotted) fractions of the Eddington
luminosity (based on the assumptions described in the text).}
\label{fig4A}
\end{figure*}

Applying the Magorrian et al. (1998) relation to the XHFS sample, we
can obtain an estimate of the underlying black hole mass, M$_{\rm
BH}$, and hence the fraction of Eddington luminosity at which any
particular XHFS nucleus is radiating.  Ho, Filippenko \& Sargent
(1997a) provide an estimate of the bulge magnitude of each galaxy in
their sample, formulated from an average, type-dependent relation of
the form discussed by Whittle (1992).  By definition all light from
ellipticals is assumed to come from the bulge (since they have no
disk); the correction to bulge luminosity from that of the full galaxy
then ranges, for example, from 1.02 mag for Sa galaxies to 2.54 mag
for Sc galaxies.  Figure~\ref{fig4A} shows a plot of the nuclear
\lx~versus L$_{\rm bulge}$ for the Seyferts and LINERs in the XHFS
sample. The diagonal lines calibrate the measurements in terms of the
fraction of the Eddington luminosity, based on the M$_{\rm BH}$ to
L$_{\rm bulge}$ relation from Magorrian et al.  Here we assume that
the X-ray emission in the \ro 0.1--2.4 keV band amounts to 15\% of the
bolometric luminosity of the AGN, which we estimate from the spectral
energy distributions of low-luminosity AGN (specifically those in
spirals) presented by Ho (1999).  The AGN in the XHFS sample would all
appear to be radiating at severely sub-Eddington luminosities, the
highest rate being $\sim 10^{-3}$ of the Eddington value, but with
$\sim 10^{-6}$ a more typical figure.

There are, of course, several factors that could induce a bias in
these measurements.  Firstly, considerable absorption in the \ro band
could easily suppress the observed luminosity by a factor $\sim 10^2$;
for example, in classical type 2 sources in which the direct line of
sight to the nucleus is completely blocked, a scattered nuclear flux
amounting to $\sim 1\%$ of the direct nuclear continuum is often
observed (\eg Turner et al. 1997).  However, many of the sources with
extremely low rates are type 1 nuclei, which in general do not exhibit
substantial soft X-ray absorption due to cold gas intrinsic to the
source (\eg George et al. 1998).  Another potential problem is that
the mass estimates (in the range $10^{6} - 10^{9} \Msun$) for the
underlying supermassive black holes powering these relatively
low-luminosity Seyfert and LINERs may be unrealistic.  The Magorrian
et al. (1998) sample is comprised predominantly of giant elliptical
and lenticular galaxies (a population which is specifically excluded
from the current work, see \S 2). Wandel (1999) has pointed out the
potential bias in the dynamical measurement towards galaxies
harbouring particularly massive black holes.  In fact, Wandel (1999)
presents results showing that black hole masses in Seyfert nuclei
(inferred via a reverberation mapping/virial mass technique) are
generally smaller by a factor $\sim 20$ than those obtained using the
Magorrian et al. relation.  Plausibly this downward scaling could be
even greater for galaxies with low-luminosity nuclei.  This is
supported by Salucci et al. (1999), who use kinematic observations of
the innermost regions of spiral galaxies to demonstrate that late-type
spirals (types Sb - Im) may contain under-massive MDOs, for a given
bulge mass, when compared to ellipticals and early-type spirals.

However, even if we apply a rather arbitrary upward scaling to our
derived Eddington fractions of, say, $10^2$ to account for the above
uncertainties, we are still left with the bulk of the XHFS AGN sample
operating in a substantially sub-Eddington regime, implying accretion
rates firmly in the range theorised for advection-dominated accretion
flows (ADAFs; Narayan \& Yi 1995; Fabian \& Rees 1995).  Colbert \&
Mushotzky (1999) discuss the ADAF scenario and an alternative
approach, namely that low-luminosity AGN found in nearby galaxies
might be powered by a new class of intermediate mass black holes
($10^{2} - 10^{4} \Msun$), accreting in their soft state.  It should
be noted, though, that this latter hypothesis does not hold in the
case of nearby galaxies where an MDO mass is known, for example NGC
4594 (Fabbiano \& Juda 1997).  Yet a further variation, suggested by
Siemiginowska et al. (1996), is that ionisation instability in the
accretion disks around super-massive black holes leads to long periods
of quiescence, during which accretion occurs at a severely
sub-Eddington rate; in this scenario LLAGNs might correspond to
systems exhibiting an extended ``low state''.

\section{Properties of the non-nuclear sample}

\begin{figure*}
\centering
\includegraphics[width=8cm]{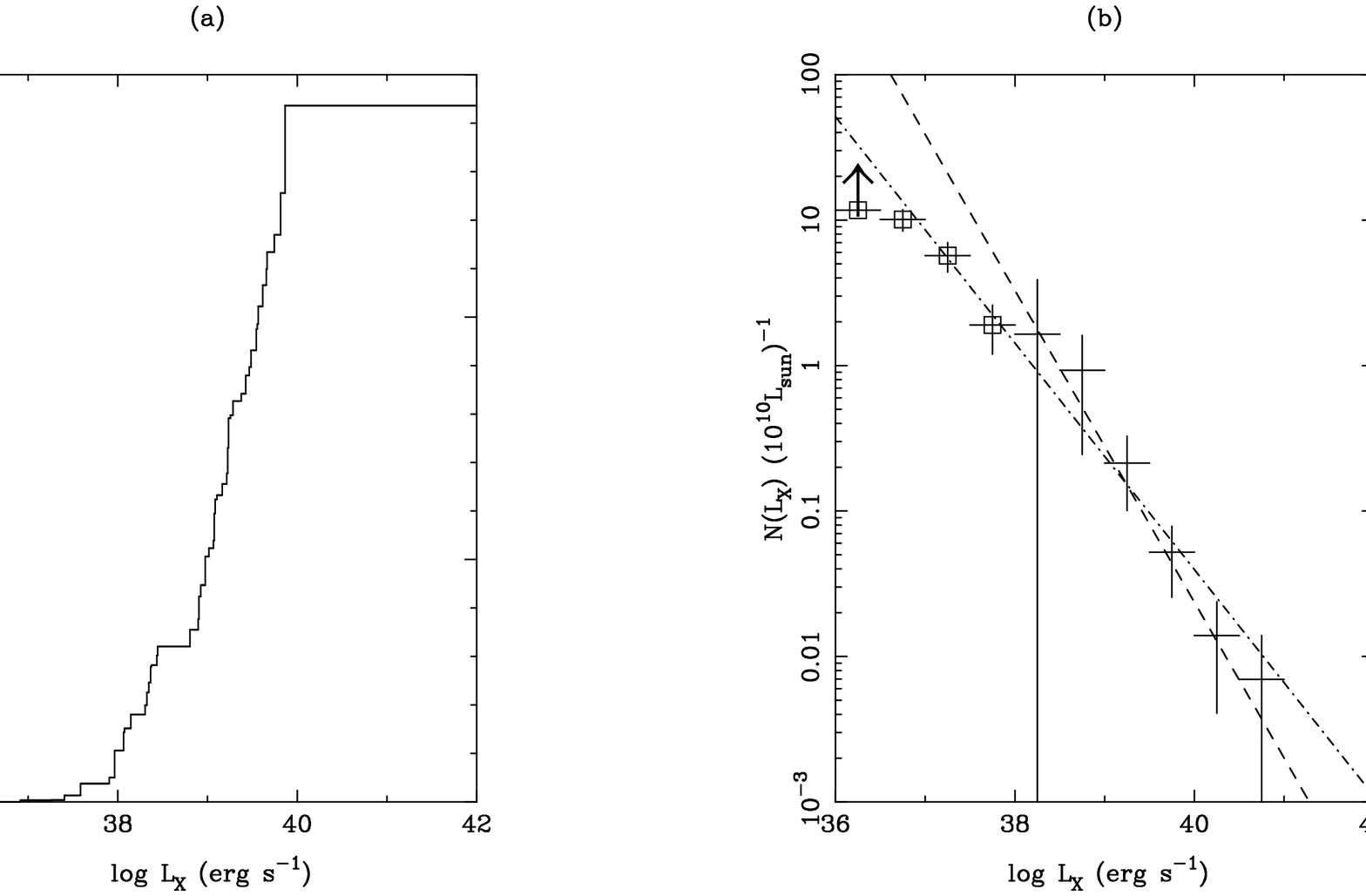}
\caption{(a) The coverage correction used in calculating the
luminosity distribution of the non-nuclear X-ray sources.  (b) The
luminosity distribution for non-nuclear sources per half decade bin in
\lx~ normalised to $10^{10} \Lsun$ in L$_B$. This distribution is
calculated for the set of 49 spiral galaxies (with $T = 0-6$ ) in
the XHFS sample.  The points (open squares) plotted below \lx$ =
10^{38} \rm~erg~s^{-1}$ are for M 31 (Supper et al. 1997).  Power-law
fits both to the XHFS data alone and combined with M31
measurements are shown as the dashed and dashed-dotted lines
respectively.}
\label{fig5}
\end{figure*}

Non-nuclear X-ray sources have been detected in 34 out of the 83
galaxies in the XHFS sample.  These sources range in luminosity from
$\sim 10^{36.5}$ to $\sim 10^{41} \ergsec$ in the 0.1--2.4 keV band as
illustrated in Figure~\ref{fig2}(a). Here we investigate the true form of 
the high luminosity end of the X-ray luminosity distribution for 
discrete sources in galaxies (excluding low-luminosity AGN).
Specifically we focus on discrete X-ray sources detected in
spiral galaxies of type SOa to Scd ($T = 0-6$), which provide the bulk
of the non-nuclear sources in our sample (120 out of 142 sources).

Since the effective sensitivity in terms of \lx ~varies from galaxy to
galaxy, it is necessary to apply a correction for the varying survey
coverage.  For each \ro HRI field we transposed the threshold flux for
point source detection (as calculated in \S 3.2) to luminosity using
the distance of the XHFS target galaxy and then constructed a
cumulative coverage curve as a function of increasing (threshold)
X-ray luminosity.  To allow for the considerable spread in absolute
magnitude across the 49 galaxies which comprise the XHFS spiral sample
(this includes the spirals with no non-nuclear source detections),
each galaxy was given a weight proportional to its optical B-band
luminosity; specifically we use a weight factor equal to L$_B$ in
units of $10^{10} \Lsun$.  The resulting coverage curve is shown in
Figure~\ref{fig5}(a). The ``true'' distribution in luminosity was
calculated by binning the observed source sample in terms of \lx~(in
half-decade logarithmic steps) and then correcting the number of
sources in each bin, $N_{i}$, for the effective coverage.  The result
was the differential luminosity distribution shown in
Figure~\ref{fig5}(b) which, in effect, represents an average per
$10^{10} \Lsun$ of optical blue luminosity. The error bars are simply
$\pm\sqrt{N_{i}}$ estimates scaled by the coverage correction.

The XHFS measurements best constrain the discrete X-ray source
population of spiral galaxies in the X-ray luminosity range
$10^{38.5}-10^{40.5} \ergsec$.  However, information is available for
X-ray sources at least two orders of magnitude less luminous, from the
extensive surveys of the local group galaxies carried out by \ro. Here
we have used the results on M31 from the \ro PSPC survey reported by
Supper et al. (1997) to extend the measurements in
Figure~\ref{fig5}(b) down to $10^{36} \ergsec$.  For this purpose we
have used the log N- log S curves for X-ray sources detected in disk
areas II, III and IV of M31 (see Figure 14, Supper et al. 1997).  We
also use the conversion 1 PSPC count s$^{-1} = 1.5 \times 10^{39}
\ergsec$ (0.1--2.4 keV) and scale the M31 source numbers to an optical
luminosity L$_{B} = 3.6 \times 10^{10} \Lsun$.  Note that the M31
measurements in the lowest bin are shown as a lower-limit since the
correction for varying sensitivity across the field of view in the
PSPC observations is a significant (and somewhat uncertain) correction
at this level of X-ray luminosity.

The measurements in Figure~\ref{fig5}(b) indicate a steep, near
power-law form for the X-ray luminosity distribution of discrete
sources in spiral galaxies.  Taking just the XHFS data points and
using a simple minimum $\chi^2$ fitting technique, we find that the
best fitting straight line in log N(\lx) - log \lx~space has a
power-law slope of $-1.08^{+0.40}_{-0.22}$. However, when the XHFS
data are combined with the M31 measurements a somewhat flatter slope
of $-0.78^{+0.07}_{-0.09}$ is obtained. These two best-fitting
power-law forms are illustrated in Figure~\ref{fig5}(b). The flatter
relation appears to be broadly consistent, within the errors, with all
the available data although, in this case, there is the hint of a turn
down above \lx~$\approx 10^{40} \ergsec$.  Unfortunately the data
quality are such that we cannot rule out the possibility that a break
in the power-law occurs at even lower luminosity (\eg at $\sim 10^{39}
\ergsec$).

The above analysis is based on the X-ray sources detected within the
confines of the optical disc of the sample of spiral galaxies but
excludes the sources located within the nominal $25''$ nuclear error
circle, some of which may well correspond to components of the
galactic source population other than low-luminosity AGN.  Thus our
luminosity distribution strictly represents a lower-limit estimate;
fortunately the comparison with M31 is on a similar basis in that the
M31 measurements exclude the bulge region, \ie a region within radius
$5' (= 1$ kpc) of the nucleus of M31.  If we include both the nuclear
X-ray sources associated with \hii and NoEL galaxies (9 sources) and
also those associated with Seyfert and LINER nuclei (a further 20
sources), the effect, as expected, is a flattening of the power-law
fit to the XHFS data to a slope of $-0.65^{+0.20}_{-0.18}$, a value
which changes only marginally when the M31 data are also included.  A
further potential bias arises, of course, due to the limited angular
resolution of the HRI. In this case the effect of closely-spaced
discrete sources blending into a single detection, is to induce a
flattening of the measured luminosity function. It is quite possible
that this latter effect partly compensates for the fact that the
nuclear sources are excluded in our primary analysis.

The best-fit straight line to the combined XHFS and M31 datasets in
Figure~\ref{fig5}(b) corresponds to a differential luminosity
distribution of the form $dN/dL_{38} = (1.0 \pm 0.2) L_{38}^{-1.8}$
sources per unit $L_{38}$ per $10^{10}$ $\Lsun$, where $L_{38}$ is the
X-ray luminosity in units of $10^{38} \ergsec$.  This form for the
luminosity distribution is in excellent agreement with that recently
reported for the X-ray source population in M101 by Wang, Immler \&
Pietsch (1999) except that their normalisation (per $10^{10} \Lsun$)
is a factor $\sim 4$ higher than our sample-averaged value\footnote{We
detect a total of 27 sources in M101 (= NGC 5457) - see Appendix A;
Wang et al. (1999) exploit an ultra-deep co-addition of all the
available HRI data and a detection threshold set at $3.5 \sigma$ in
deriving a list of 51 X-ray sources, about half of which are probably
associated with M101.}.  Almost certainly this represents a real
over-density of bright X-ray sources in M101 with respect to the norm
for spiral galaxies and may be related to a number of factors such as
the relatively low metallicity in the outer disk of M101 and the
presence across the galaxy of numerous giant HII clouds and vigorous
star-forming regions (Trinchieri, Fabbiano \& Romaine 1990; Wang et
al. 1999).

Integration of the luminosity distribution derived above for \lx~ in
the range $10^{36} - 10^{40} \ergsec$, leads to an \lx/L$_{B}$ ratio
for the discrete X-ray source population in spiral galaxies of $1.1
\times 10^{39} \ergsec (10^{10} \Lsun)^{-1}$.  In a recent paper Brown
\& Bregman (1998) have used \ro data to study the \lx~versus L$_{B}$
relation for a complete sample of early type galaxies. These authors
estimate the contribution to the soft X-ray emission from stellar
X-ray sources on the basis of X-ray spectral measurements for Cen A
(the nearest non-dwarf elliptical galaxy).  Specifically they use the
results of Turner et al. (1997) who identify a hard spectral component
in Cen A with stellar sources but attribute a soft component ($kT
\approx 0.3$ keV) to emission from diffuse hot gas. The hard spectral
component implies an \lx/L$_{B}$ ratio for Cen A of $1.0 \times
10^{39} \ergsec (10^{10} \Lsun)^{-1}$, in excellent agreement with the
value we derived above for spiral galaxies. (Note we have scaled up
the hard X-ray luminosity for Cen A quoted by Brown \& Bregman (1998)
for the 0.5--2 keV band by a factor of 2.3 to account for our wider
0.1--2.4 keV bandpass).

The luminosity function defined above may also be integrated to give
an estimate of the frequency of occurrence in spiral galaxies of
``super-luminous'' galactic X-ray sources. Here we define a
super-luminous source (SLS) to be one with a luminosity above \lx~ $=
10^{38.3} \ergsec$, which is the Eddington limit for accretion onto a
$1.4 \Msun$ neutron star.  We find that, on average, there are $\sim
0.7$ such sources per $10^{10} \Lsun$. The source catalogue in
Appendix A actually lists 85 non-nuclear SLS, 28 of which have \lx~in
excess of $10^{39.0} \ergsec$.

The nature of the SLS population in nearby galaxies has been the
subject of considerable debate following the discovery of many such
objects in \ein and, more recently, \ro images (e.g. Fabbiano 1989;
1995; Read et al. 1997). Populations of galactic X-ray source which
may generate X-ray luminosities in excess of $10^{38.3} \ergsec$
include XRBs powered by accretion onto a stellar mass ($M_{BH} <
10{^2} \Msun$) black hole and very young SNRs in which the supernova
ejecta interact strongly with the circumstellar medium (e.g. Fabbiano,
1996 and references therein).  Recently it has also been suggested
that very luminous remnants observed in M101 may be the relics of
``hypernovae'', that is to say local manifestations of gamma-ray burst
activity (Wang 1999).

An important step when dealing with a point-like source which is
apparently super-luminous is to consider whether it represents one
discrete object or a complex of sources (for example, a dense grouping
of high and/or low mass XRBs which individually would not be
categorised as a SLS population). In this respect large amplitude
variability provides a pointer towards the presence of either a single
compact source or a grouping of no more than a few such objects.  We
have used the multiple HRI observations, where available, to identify
those sources exhibiting significant (\ie factor $ \ge 2$) variability
(see Appendix A). It turns out that only 1 out 19 of the non-nuclear
SLS sources is flagged as variable, but unfortunately due to the
limited temporal sampling it is difficult to draw any firm conclusions
from this result.

In conclusion it seems likely that the non-nuclear SLS sources in the
XHFS sample represent a heterogeneous mix of luminous stellar mass
black hole XRBs, complexes of near-Eddington limited (or perhaps
super-Eddington) high- and low-mass XRBs and a number of examples of
recent supernovae and young SNRs. However, the detailed follow-up of
individual sources is beyond the scope of the present paper.

\section{Summary}

This paper utilises the substantial resources of both the \ro HRI
public archive and the HFS sample to study the statistical properties
of discrete X-ray sources in bright nearby galaxies.  Our main
findings may be summarised as follows:

\begin{description}
\item[(i)] 
Our XHFS catalogue lists 187 discrete X-ray sources detected within
the optical extent of 83 nearby galaxies.  The 45 sources which lie
within $25''$ of the optical nucleus of the host galaxy are designated
as ``nuclear'' X-ray sources with the remainder comprising our
``non-nuclear'' source sample.
\item[(ii)]
Contamination of the XHFS sample by background and/or foreground
objects is low.  It amounts to less than 1 in 8 of the non-nuclear
sample and is totally neglible in the nuclear sample.
\item[(iii)]
The detection rate of nuclear X-ray sources and the average X-ray
luminosity is substantially higher for those galaxies potentially
hosting a low-luminosity AGN (Seyferts and LINERs) compared to the
galaxies exhibiting only low-excitation emission spectra (the \hii
galaxies) or no emission lines.
\item[(iv)]
The \lx/L$_{\rm H\alpha}$ ratio of the putative AGN in the XHFS sample
is in the same range as that observed for higher luminosity objects,
such as QSOs, but extends two magnitudes fainter in luminosity than
previously reported. Using the M$_{\rm BH}$ to L$_{\rm bulge}$
relation from Magorrian et al. (1998), we find that the low-luminosity
AGN in the XHFS sample are radiating at severely sub-Eddington rates
(\ie $\sim 10^{-6}$ -- 10$^{-3}$ of the Eddington luminosity).
\item[(v)]
The luminosity distribution of the non-nuclear sources in spiral
galaxies of type SOa - Scd is $dN/dL_{38} = (1.0 \pm 0.2)
L_{38}^{-1.8}$ sources per unit $L_{38}$ per $10^{10} \Lsun$ in the
$10^{36} - 10^{40} \ergsec$ range.  This integrates to give a
\lx/L$_B$ ratio of $1.1 \times 10^{39} \ergsec (10^{10} \Lsun)^{-1}$.
\item[(vi)]
The median luminosity of the non-nuclear sources is $\sim 10$ times
lower than that of nuclear sources.  However, it is still in excess of
the Eddington luminosity for a 1.4 solar mass neutron star ($\approx
10^{38.3} \ergsec$), implying a substantial population of
super-luminous X-ray sources in nearby galaxies. Our estimate of the
incidence such sources is $\sim 0.7$ per $10^{10} \Lsun$.

\end{description}

Although we discuss the possible nature of the super-luminous X-ray
sources revealed in the XHFS survey in large numbers, we have not
pursued the identification of individual sources, and indeed such a
programme is greatly hampered by the limitations of the current X-ray
data.  This should change with the advent of the new generation of
X-ray observatories, most particularly \xmm and {\it CHANDRA\/}, whose
combination of excellent spatial resolution, wide bandpass medium
resolution spectroscopy and high throughput are ideally suited to
detailed studies of the bright X-ray source populations in nearby
galaxies.

\vspace{1cm}

{\noindent \bf ACKNOWLEDGMENTS}

\vspace{2mm}

TPR gratefully acknowledges financial support from PPARC.  The X-ray
data used in this work were all obtained from the Leicester Database
and Archive Service (LEDAS) at the Department of Physics \& Astronomy,
University of Leicester.  The Digitised Sky Survey was produced at the
Space Telescope Science Institute, under US government grant NAG
W-2166 from the original National Geographic--Palomar Sky Survey
plates. We are grateful to the anonymous referee for suggesting
a number of improvements to this paper.

\newpage

\appendix

\section{CATALOGUE OF DETECTED X-RAY SOURCES}

The complete, compressed catalogue of individual X-ray sources
observed within the \d25 ellipse of the XHFS galaxies is presented in
Table~\ref{apptab1}.  All position and count rate data are determined
by the {\small PSS} algorithm, and the nuclear separation is the
observed distance between the HFS nuclear position and the X-ray
source.  Where there is more than one detection of a source, the data
are combined with a weighting commensurate with the observation durations. 
The time-averaged count rates have been converted to flux as 
described in \S 2.  Finally, the notes column shows an ``N'' if the source 
is a nuclear X-ray source, and a ``v'' if it is designated a variable
source on the basis of its count rate varying by a factor $\ge 2$
between detections in two separate observation epochs.

\begin{table*}
\centering
\caption{Catalogue of the 187 discrete X-ray sources which comprise the XHFS 
sample.}
\label{apptab1}
\begin{tabular}{llcccccc}\hline
Source designation	& Observations	& RA(2000)	& dec(2000)	& Count rate	& Nuclear offset	& log \lx	& Notes \\
			&		&		&		& ($\times 10^{-3} \ctsec$)	& ($''$)	& ($\ergsec$)	& \\\hline
IC 10    X-1  & AB      & 00$^h$20$^m$29.7$^s$ & 59$^{\circ}$16$'$48$"$ &   8.7 $\pm$ 0.4 &  57.5 & 38.41 & \\
NGC 185  X-1  &         & 00$^h$39$^m$02.7$^s$ & 48$^{\circ}$24$'$09$"$ &   1.7 $\pm$ 0.4 & 243.5 & 36.87 & \\
NGC 205  X-1  &         & 00$^h$39$^m$52.5$^s$ & 41$^{\circ}$45$'$17$"$ &   1.1 $\pm$ 0.3 & 405.4 & 36.59 & \\
NGC 205  X-2  &         & 00$^h$40$^m$08.4$^s$ & 41$^{\circ}$40$'$12$"$ &   1.2 $\pm$ 0.3 & 167.5 & 36.64 & \\
NGC 221  X-1  &         & 00$^h$42$^m$35.6$^s$ & 40$^{\circ}$48$'$32$"$ &   2.7 $\pm$ 0.6 & 224.7 & 36.99 & \\
NGC 221  X-2  &         & 00$^h$42$^m$43.2$^s$ & 40$^{\circ}$51$'$47$"$ &  13.4 $\pm$ 1.1 &  12.2 & 37.69 & N \\
NGC 404  X-1  &         & 01$^h$09$^m$26.9$^s$ & 35$^{\circ}$43$'$01$"$ &   1.2 $\pm$ 0.3 &   4.0 & 37.68 & N \\
NGC 891  X-1  &         & 02$^h$22$^m$26.6$^s$ & 42$^{\circ}$18$'$27$"$ &   0.6 $\pm$ 0.1 & 167.2 & 38.63 & \\
NGC 891  X-2  &         & 02$^h$22$^m$30.8$^s$ & 42$^{\circ}$19$'$07$"$ &   0.6 $\pm$ 0.1 & 110.6 & 38.60 & \\
NGC 891  X-3  &         & 02$^h$22$^m$31.5$^s$ & 42$^{\circ}$19$'$54$"$ &   4.8 $\pm$ 0.3 &  64.4 & 39.53 & \\
NGC 891  X-4  &         & 02$^h$22$^m$31.6$^s$ & 42$^{\circ}$20$'$19$"$ &   1.9 $\pm$ 0.2 &  43.6 & 39.12 & \\
NGC 891  X-5  &         & 02$^h$22$^m$45.7$^s$ & 42$^{\circ}$25$'$53$"$ &   1.6 $\pm$ 0.2 & 327.7 & 39.06 & \\
IC 342   X-1  &         & 03$^h$45$^m$56.0$^s$ & 68$^{\circ}$04$'$55$"$ &   8.2 $\pm$ 0.7 & 303.6 & 39.00 & \\
IC 342   X-2  &         & 03$^h$46$^m$07.3$^s$ & 68$^{\circ}$07$'$06$"$ &   1.9 $\pm$ 0.4 & 250.2 & 38.36 & \\
IC 342   X-3  &         & 03$^h$46$^m$16.6$^s$ & 68$^{\circ}$11$'$14$"$ &   2.7 $\pm$ 0.4 & 378.9 & 38.52 & \\
IC 342   X-4  &         & 03$^h$46$^m$46.4$^s$ & 68$^{\circ}$09$'$47$"$ &   4.9 $\pm$ 0.6 & 245.0 & 38.78 & \\
IC 342   X-5  &         & 03$^h$46$^m$49.6$^s$ & 68$^{\circ}$05$'$47$"$ &   7.4 $\pm$ 0.7 &   3.9 & 38.96 & N \\
IC 342   X-6  &         & 03$^h$46$^m$58.5$^s$ & 68$^{\circ}$06$'$18$"$ &   3.8 $\pm$ 0.5 &  61.5 & 38.67 & \\
IC 342   X-7  &         & 03$^h$48$^m$07.6$^s$ & 68$^{\circ}$04$'$54$"$ &   1.3 $\pm$ 0.3 & 439.6 & 38.19 & \\
NGC 1569 X-1  &         & 04$^h$30$^m$48.7$^s$ & 64$^{\circ}$50$'$47$"$ &   5.3 $\pm$ 0.8 &  10.3 & 38.22 & N \\
NGC 1961 X-1  &         & 05$^h$42$^m$04.4$^s$ & 69$^{\circ}$22$'$43$"$ &   1.6 $\pm$ 0.2 &   7.4 & 40.56 & N \\
NGC 2276 X-1  & AB      & 07$^h$26$^m$49.5$^s$ & 85$^{\circ}$45$'$49$"$ &   1.5 $\pm$ 0.2 &  47.9 & 40.16 & \\
NGC 2403 X-1  &         & 07$^h$36$^m$26.5$^s$ & 65$^{\circ}$35$'$37$"$ &  13.1 $\pm$ 0.8 & 161.2 & 39.19 & \\
NGC 2403 X-2  &         & 07$^h$36$^m$52.9$^s$ & 65$^{\circ}$34$'$54$"$ &   1.0 $\pm$ 0.3 &  68.0 & 38.09 & \\
NGC 2403 X-3  &         & 07$^h$36$^m$56.8$^s$ & 65$^{\circ}$35$'$38$"$ &   4.0 $\pm$ 0.5 &  37.2 & 38.67 & \\
NGC 2403 X-4  &         & 07$^h$37$^m$03.4$^s$ & 65$^{\circ}$39$'$33$"$ &   3.6 $\pm$ 0.4 & 222.2 & 38.63 & \\
NGC 2782 X-1  &         & 09$^h$14$^m$04.4$^s$ & 40$^{\circ}$07$'$38$"$ &   1.7 $\pm$ 0.4 &  49.4 & 40.11 & \\
NGC 2782 X-2  &         & 09$^h$14$^m$05.4$^s$ & 40$^{\circ}$06$'$46$"$ &   7.3 $\pm$ 0.7 &  12.4 & 40.75 & N \\
NGC 2903 X-1  &         & 09$^h$32$^m$05.6$^s$ & 21$^{\circ}$32$'$32$"$ &   1.0 $\pm$ 0.3 & 169.4 & 38.39 & \\
NGC 2903 X-2  &         & 09$^h$32$^m$09.9$^s$ & 21$^{\circ}$31$'$01$"$ &   1.6 $\pm$ 0.4 &  65.1 & 38.60 & \\
NGC 2903 X-3  &         & 09$^h$32$^m$10.4$^s$ & 21$^{\circ}$30$'$05$"$ &   9.6 $\pm$ 1.0 &   9.1 & 39.38 & N \\
NGC 2976 X-1  & ABC     & 09$^h$47$^m$14.5$^s$ & 67$^{\circ}$55$'$42$"$ &   2.3 $\pm$ 0.2 &  41.4 & 37.84 & v \\
NGC 3031 X-1  & F       & 09$^h$54$^m$51.9$^s$ & 69$^{\circ}$02$'$48$"$ &   3.7 $\pm$ 0.5 & 251.6 & 37.69 & \\
NGC 3031 X-2  & ACDEFG  & 09$^h$55$^m$01.0$^s$ & 69$^{\circ}$07$'$41$"$ &   1.3 $\pm$ 0.1 & 287.4 & 37.25 & \\
NGC 3031 X-3  & ABCDFG  & 09$^h$55$^m$03.0$^s$ & 68$^{\circ}$56$'$15$"$ &   1.2 $\pm$ 0.1 & 504.0 & 37.21 & v \\
NGC 3031 X-4  & A       & 09$^h$55$^m$10.8$^s$ & 69$^{\circ}$04$'$03$"$ &   4.5 $\pm$ 0.5 & 137.4 & 37.78 & \\
NGC 3031 X-5  & ABCDEFG & 09$^h$55$^m$11.3$^s$ & 69$^{\circ}$04$'$58$"$ &   1.8 $\pm$ 0.1 & 144.8 & 37.38 & v \\
NGC 3031 X-6  & F       & 09$^h$55$^m$13.2$^s$ & 69$^{\circ}$08$'$25$"$ &   1.8 $\pm$ 0.4 & 287.3 & 37.37 & \\
NGC 3031 X-7  & ABCDEFG & 09$^h$55$^m$23.1$^s$ & 69$^{\circ}$05$'$07$"$ &   1.4 $\pm$ 0.1 &  94.4 & 37.26 & \\
NGC 3031 X-8  & ABCDEFG & 09$^h$55$^m$25.3$^s$ & 69$^{\circ}$09$'$56$"$ &   4.5 $\pm$ 0.2 & 355.0 & 37.77 & \\
NGC 3031 X-9  & BCDEFG  & 09$^h$55$^m$25.7$^s$ & 69$^{\circ}$01$'$08$"$ &  12.5 $\pm$ 0.4 & 186.4 & 38.22 & v \\
NGC 3031 X-10 & F       & 09$^h$55$^m$27.2$^s$ & 69$^{\circ}$02$'$50$"$ &   1.3 $\pm$ 0.4 &  90.2 & 37.24 & \\
NGC 3031 X-11 & ABCDEFG & 09$^h$55$^m$34.0$^s$ & 69$^{\circ}$00$'$28$"$ &  25.7 $\pm$ 0.5 & 217.9 & 38.53 & \\
NGC 3031 X-12 & C       & 09$^h$55$^m$34.1$^s$ & 69$^{\circ}$02$'$42$"$ &   1.3 $\pm$ 0.4 &  84.9 & 37.24 & \\
NGC 3031 X-13 & ABCDEFG & 09$^h$55$^m$34.2$^s$ & 69$^{\circ}$03$'$50$"$ & 334.0 $\pm$ 1.6 &  19.6 & 39.64 & N,v \\
NGC 3031 X-14 & CF      & 09$^h$55$^m$38.5$^s$ & 69$^{\circ}$02$'$40$"$ &   1.9 $\pm$ 0.3 &  86.6 & 37.41 & \\
NGC 3031 X-15 & BF      & 09$^h$55$^m$50.2$^s$ & 68$^{\circ}$58$'$33$"$ &   1.1 $\pm$ 0.2 & 340.4 & 37.16 & \\
NGC 3031 X-16 & ABCDEFG & 09$^h$55$^m$51.0$^s$ & 69$^{\circ}$05$'$28$"$ &   2.8 $\pm$ 0.2 & 113.8 & 37.56 & \\
NGC 3031 X-17 & ABC     & 09$^h$56$^m$10.3$^s$ & 69$^{\circ}$01$'$01$"$ &   1.4 $\pm$ 0.2 & 258.7 & 37.25 & v \\
NGC 3031 X-18 & ABCDG   & 09$^h$56$^m$15.4$^s$ & 68$^{\circ}$57$'$18$"$ &   1.5 $\pm$ 0.1 & 458.3 & 37.29 & v \\
NGC 3031 X-19 & BD      & 09$^h$56$^m$37.7$^s$ & 69$^{\circ}$00$'$25$"$ &   1.1 $\pm$ 0.2 & 396.0 & 37.18 & \\
NGC 3031 X-20 & BC      & 09$^h$57$^m$02.5$^s$ & 68$^{\circ}$54$'$53$"$ &   1.2 $\pm$ 0.2 & 721.1 & 37.20 & \\
NGC 3079 X-1  & ABCDE   & 10$^h$01$^m$59.0$^s$ & 55$^{\circ}$40$'$45$"$ &   2.6 $\pm$ 0.2 &  14.0 & 39.73 & N,v \\
NGC 3147 X-1  & ABC     & 10$^h$16$^m$54.4$^s$ & 73$^{\circ}$24$'$00$"$ &  53.2 $\pm$ 0.9 &   3.7 & 41.74 & N \\
NGC 3190 X-1  &         & 10$^h$18$^m$05.8$^s$ & 21$^{\circ}$49$'$59$"$ &   2.1 $\pm$ 0.3 &  21.1 & 39.79 & N \\
NGC 3226 X-1  &         & 10$^h$23$^m$27.3$^s$ & 19$^{\circ}$53$'$54$"$ &   3.7 $\pm$ 0.4 &   8.0 & 40.07 & N \\
NGC 3227 X-1  &         & 10$^h$23$^m$30.8$^s$ & 19$^{\circ}$51$'$51$"$ &  84.6 $\pm$ 1.8 &   4.8 & 41.32 & N \\
NGC 3310 X-1  &         & 10$^h$38$^m$46.1$^s$ & 53$^{\circ}$30$'$11$"$ &  23.5 $\pm$ 0.9 &  47.9 & 40.62 & \\
NGC 3310 X-2  &         & 10$^h$38$^m$49.8$^s$ & 53$^{\circ}$29$'$33$"$ &   1.0 $\pm$ 0.2 &  40.6 & 39.25 & \\
NGC 3377 X-1  &         & 10$^h$47$^m$41.7$^s$ & 13$^{\circ}$58$'$26$"$ &   0.8 $\pm$ 0.2 &  47.5 & 38.53 & \\
NGC 3377 X-2  &         & 10$^h$47$^m$42.7$^s$ & 13$^{\circ}$59$'$07$"$ &   1.2 $\pm$ 0.3 &   3.9 & 38.68 & N \\
\end{tabular}
\end{table*}

\begin{table*}
\centering
\begin{tabular}{llcccccc}\hline
Source designation	& Observations	& RA(2000)	& dec(2000)	& Count rate	& Nuclear offset	& log \lx	& Notes \\
			&		&		&		& ($\times 10^{-3} \ctsec$)	& ($''$)	& ($\ergsec$)	& \\\hline
NGC 3379 X-1  &         & 10$^h$47$^m$50.2$^s$ & 12$^{\circ}$34$'$55$"$ &   5.0 $\pm$ 0.5 &   5.4 & 39.30 & N \\
NGC 3395 X-1  &         & 10$^h$49$^m$50.3$^s$ & 32$^{\circ}$59$'$07$"$ &   1.4 $\pm$ 0.3 &  26.6 & 39.75 & \\
NGC 3628 X-1  &         & 11$^h$20$^m$15.7$^s$ & 13$^{\circ}$35$'$23$"$ &  12.3 $\pm$ 1.0 &  31.9 & 39.62 & \\
NGC 3628 X-2  &         & 11$^h$20$^m$37.3$^s$ & 13$^{\circ}$34$'$38$"$ &   3.3 $\pm$ 0.6 & 296.1 & 39.04 & \\
NGC 3998 X-1  &         & 11$^h$57$^m$50.4$^s$ & 55$^{\circ}$28$'$31$"$ &   1.8 $\pm$ 0.5 &  78.3 & 39.63 & \\
NGC 3998 X-2  &         & 11$^h$57$^m$56.4$^s$ & 55$^{\circ}$27$'$10$"$ & 161.7 $\pm$ 3.6 &  17.4 & 41.59 & N \\
NGC 4051 X-1  &         & 12$^h$03$^m$09.8$^s$ & 44$^{\circ}$31$'$54$"$ &1168.9 $\pm$10.9 &  12.9 & 42.25 & N \\
NGC 4088 X-1  &         & 12$^h$05$^m$32.8$^s$ & 50$^{\circ}$32$'$43$"$ &   2.5 $\pm$ 0.6 &  23.2 & 39.62 & N \\
NGC 4150 X-1  &         & 12$^h$10$^m$35.0$^s$ & 30$^{\circ}$23$'$56$"$ &  25.3 $\pm$ 1.7 &  24.1 & 40.11 & N \\
NGC 4151 X-1  & A       & 12$^h$10$^m$22.4$^s$ & 39$^{\circ}$21$'$46$"$ &   0.5 $\pm$ 0.1 & 185.6 & 39.03 & \\
NGC 4151 X-2  & AB      & 12$^h$10$^m$22.8$^s$ & 39$^{\circ}$23$'$12$"$ &   0.6 $\pm$ 0.1 & 124.7 & 39.15 & \\
NGC 4151 X-3  & AB      & 12$^h$10$^m$33.0$^s$ & 39$^{\circ}$24$'$18$"$ & 293.0 $\pm$ 1.2 &  10.7 & 41.84 & N \\
NGC 4203 X-1  &         & 12$^h$15$^m$05.3$^s$ & 33$^{\circ}$11$'$49$"$ &  59.0 $\pm$ 1.6 &  15.7 & 40.46 & N \\
NGC 4214 X-1  &         & 12$^h$15$^m$38.3$^s$ & 36$^{\circ}$19$'$17$"$ &   2.7 $\pm$ 0.3 &  37.7 & 38.26 & \\
NGC 4235 X-1  &         & 12$^h$17$^m$10.3$^s$ & 07$^{\circ}$11$'$25$"$ & 122.1 $\pm$ 2.6 &   7.6 & 41.91 & N \\
NGC 4258 X-1  & B       & 12$^h$18$^m$44.6$^s$ & 47$^{\circ}$17$'$21$"$ &   1.0 $\pm$ 0.3 & 147.9 & 38.40 & \\
NGC 4258 X-2  & AB      & 12$^h$18$^m$45.9$^s$ & 47$^{\circ}$18$'$26$"$ &   1.3 $\pm$ 0.2 & 121.1 & 38.49 & \\
NGC 4258 X-3  & AB      & 12$^h$18$^m$45.8$^s$ & 47$^{\circ}$24$'$18$"$ &   1.2 $\pm$ 0.2 & 376.7 & 38.46 & \\
NGC 4258 X-4  & AB      & 12$^h$18$^m$55.2$^s$ & 47$^{\circ}$16$'$45$"$ &   1.4 $\pm$ 0.2 & 100.4 & 38.52 & \\
NGC 4258 X-5  & B       & 12$^h$18$^m$57.1$^s$ & 47$^{\circ}$21$'$20$"$ &   1.2 $\pm$ 0.3 & 178.7 & 38.46 & \\
NGC 4258 X-6  & AB      & 12$^h$18$^m$58.1$^s$ & 47$^{\circ}$18$'$03$"$ &   9.1 $\pm$ 0.5 &  18.7 & 39.35 & N \\
NGC 4258 X-7  & AB      & 12$^h$18$^m$58.3$^s$ & 47$^{\circ}$16$'$01$"$ &   1.5 $\pm$ 0.2 & 140.1 & 38.57 & \\
NGC 4258 X-8  & AB      & 12$^h$19$^m$04.6$^s$ & 47$^{\circ}$18$'$27$"$ &   2.4 $\pm$ 0.3 &  69.4 & 38.76 & \\
NGC 4291 X-1  & AB      & 12$^h$20$^m$18.2$^s$ & 75$^{\circ}$22$'$09$"$ &   3.0 $\pm$ 0.3 &   6.8 & 40.22 & N \\
NGC 4321 X-1  &         & 12$^h$22$^m$51.1$^s$ & 15$^{\circ}$49$'$33$"$ &   0.9 $\pm$ 0.2 &  59.2 & 39.17 & \\
NGC 4321 X-2  &         & 12$^h$22$^m$55.0$^s$ & 15$^{\circ}$49$'$16$"$ &   8.2 $\pm$ 0.5 &   8.2 & 40.13 & N \\
NGC 4321 X-3  &         & 12$^h$22$^m$58.6$^s$ & 15$^{\circ}$49$'$07$"$ &   0.7 $\pm$ 0.2 &  52.6 & 39.09 & \\
NGC 4321 X-4  &         & 12$^h$22$^m$58.9$^s$ & 15$^{\circ}$47$'$49$"$ &   0.8 $\pm$ 0.2 & 109.0 & 39.14 & \\
NGC 4388 X-1  &         & 12$^h$25$^m$47.3$^s$ & 12$^{\circ}$39$'$38$"$ &   6.3 $\pm$ 0.9 &   5.1 & 40.03 & N \\
NGC 4395 X-1  &         & 12$^h$26$^m$01.9$^s$ & 33$^{\circ}$31$'$31$"$ &  16.8 $\pm$ 1.3 & 177.9 & 39.06 & \\
NGC 4435 X-1  &         & 12$^h$27$^m$40.9$^s$ & 13$^{\circ}$04$'$44$"$ &   1.9 $\pm$ 0.4 &   8.2 & 39.51 & N \\
NGC 4438 X-1  &         & 12$^h$27$^m$46.0$^s$ & 13$^{\circ}$00$'$26$"$ &   8.2 $\pm$ 0.7 &   9.4 & 40.15 & N \\
NGC 4449 X-1  & ABC     & 12$^h$28$^m$09.7$^s$ & 44$^{\circ}$05$'$04$"$ &   6.4 $\pm$ 0.4 &  61.9 & 38.48 & \\
NGC 4449 X-2  & BC      & 12$^h$28$^m$10.5$^s$ & 44$^{\circ}$05$'$51$"$ &   2.0 $\pm$ 0.3 &  40.0 & 37.97 & \\
NGC 4449 X-3  & ABC     & 12$^h$28$^m$11.3$^s$ & 44$^{\circ}$03$'$33$"$ &   2.9 $\pm$ 0.3 & 134.8 & 38.14 & v \\
NGC 4449 X-4  & ABC     & 12$^h$28$^m$11.6$^s$ & 44$^{\circ}$06$'$41$"$ &   5.5 $\pm$ 0.4 &  63.7 & 38.41 & \\
NGC 4449 X-5  & B       & 12$^h$28$^m$13.3$^s$ & 44$^{\circ}$06$'$22$"$ &   5.3 $\pm$ 0.6 &  39.3 & 38.40 & \\
NGC 4449 X-6  & AB      & 12$^h$28$^m$15.0$^s$ & 44$^{\circ}$05$'$42$"$ &   1.4 $\pm$ 0.3 &  10.0 & 37.80 & N \\
NGC 4449 X-7  & ABC     & 12$^h$28$^m$18.4$^s$ & 44$^{\circ}$06$'$29$"$ &   3.3 $\pm$ 0.3 &  64.8 & 38.19 & v \\
NGC 4485 X-1  & AB      & 12$^h$30$^m$30.9$^s$ & 41$^{\circ}$41$'$39$"$ &   2.8 $\pm$ 0.3 &  18.9 & 39.12 & N \\
NGC 4490 X-1  & AB      & 12$^h$30$^m$32.4$^s$ & 41$^{\circ}$39$'$12$"$ &   3.4 $\pm$ 0.3 &  67.2 & 39.06 & \\
NGC 4490 X-2  & AB      & 12$^h$30$^m$36.7$^s$ & 41$^{\circ}$38$'$34$"$ &   6.1 $\pm$ 0.4 &  13.9 & 39.31 & N \\
NGC 4490 X-3  & A       & 12$^h$30$^m$38.8$^s$ & 41$^{\circ}$37$'$43$"$ &   1.4 $\pm$ 0.3 &  47.0 & 38.66 & \\
NGC 4490 X-4  & AB      & 12$^h$30$^m$43.6$^s$ & 41$^{\circ}$38$'$14$"$ &   2.4 $\pm$ 0.3 &  81.9 & 38.91 & \\
NGC 4559 X-1  &         & 12$^h$35$^m$52.0$^s$ & 27$^{\circ}$56$'$01$"$ &  17.5 $\pm$ 0.6 & 123.7 & 39.94 & \\
NGC 4559 X-2  &         & 12$^h$35$^m$56.5$^s$ & 27$^{\circ}$55$'$26$"$ &   0.8 $\pm$ 0.2 & 131.7 & 38.62 & \\
NGC 4559 X-3  &         & 12$^h$35$^m$56.7$^s$ & 27$^{\circ}$59$'$23$"$ &   0.8 $\pm$ 0.2 & 108.1 & 38.61 & \\
NGC 4559 X-4  &         & 12$^h$35$^m$58.9$^s$ & 27$^{\circ}$57$'$40$"$ &  11.3 $\pm$ 0.5 &  13.1 & 39.75 & N \\
NGC 4559 X-5  &         & 12$^h$36$^m$03.5$^s$ & 27$^{\circ}$57$'$55$"$ &   0.9 $\pm$ 0.2 &  75.9 & 38.67 & \\
NGC 4569 X-1  &         & 12$^h$36$^m$50.1$^s$ & 13$^{\circ}$09$'$43$"$ &   5.1 $\pm$ 0.6 &  21.5 & 39.93 & N \\
NGC 4631 X-1  & AB      & 12$^h$41$^m$55.7$^s$ & 32$^{\circ}$32$'$12$"$ &   5.2 $\pm$ 0.5 & 147.7 & 39.11 & \\
NGC 4631 X-2  & A       & 12$^h$41$^m$57.5$^s$ & 32$^{\circ}$31$'$59$"$ &   2.0 $\pm$ 0.5 & 127.1 & 38.69 & \\
NGC 4651 X-1  &         & 12$^h$43$^m$42.8$^s$ & 16$^{\circ}$23$'$34$"$ &   1.3 $\pm$ 0.3 &  15.1 & 39.31 & N \\
NGC 4656 X-1  &         & 12$^h$43$^m$41.4$^s$ & 32$^{\circ}$04$'$55$"$ &   3.0 $\pm$ 0.4 & 379.5 & 38.91 & \\
NGC 4736 X-1  & A       & 12$^h$50$^m$44.5$^s$ & 41$^{\circ}$04$'$49$"$ &   0.6 $\pm$ 0.1 & 183.4 & 37.73 & \\
NGC 4736 X-2  & A       & 12$^h$50$^m$47.8$^s$ & 41$^{\circ}$05$'$03$"$ &   0.6 $\pm$ 0.1 & 152.0 & 37.77 & \\
NGC 4736 X-3  & A       & 12$^h$50$^m$52.8$^s$ & 41$^{\circ}$02$'$47$"$ &   0.4 $\pm$ 0.1 & 272.1 & 37.62 & \\
NGC 4736 X-4  & AB      & 12$^h$50$^m$53.3$^s$ & 41$^{\circ}$07$'$08$"$ &  50.5 $\pm$ 0.7 &  12.7 & 39.69 & N \\
NGC 4736 X-5  & B       & 12$^h$51$^m$00.0$^s$ & 41$^{\circ}$11$'$00$"$ &   1.7 $\pm$ 0.3 & 231.4 & 38.23 & \\
NGC 4826 X-1  &         & 12$^h$56$^m$44.0$^s$ & 21$^{\circ}$40$'$59$"$ &   6.5 $\pm$ 1.0 &   9.6 & 38.81 & N \\
NGC 5005 X-1  &         & 13$^h$10$^m$56.6$^s$ & 37$^{\circ}$03$'$24$"$ &  13.0 $\pm$ 0.8 &   6.9 & 40.47 & N \\
\end{tabular}
\end{table*}

\begin{table*}
\centering
\begin{tabular}{llcccccc}\hline
Source designation	& Observations	& RA(2000)	& dec(2000)	& Count rate	& Nuclear offset	& log \lx	& Notes \\
			&		&		&		& ($\times 10^{-3} \ctsec$)	& ($''$)	& ($\ergsec$)	& \\\hline
NGC 5055 X-1  &         & 13$^h$15$^m$18.3$^s$ & 42$^{\circ}$03$'$57$"$ &   1.4 $\pm$ 0.4 & 373.6 & 38.59 & \\
NGC 5055 X-2  &         & 13$^h$15$^m$19.8$^s$ & 42$^{\circ}$02$'$55$"$ &  16.5 $\pm$ 1.3 & 340.8 & 39.65 & \\
NGC 5055 X-3  &         & 13$^h$15$^m$30.5$^s$ & 42$^{\circ}$03$'$07$"$ &   2.2 $\pm$ 0.5 & 229.4 & 38.78 & \\
NGC 5055 X-4  &         & 13$^h$15$^m$41.0$^s$ & 42$^{\circ}$01$'$42$"$ &   2.8 $\pm$ 0.6 &  98.1 & 38.88 & \\
NGC 5055 X-5  &         & 13$^h$15$^m$47.0$^s$ & 42$^{\circ}$01$'$52$"$ &   3.7 $\pm$ 0.7 &  31.3 & 39.01 & \\
NGC 5055 X-6  &         & 13$^h$15$^m$47.1$^s$ & 42$^{\circ}$00$'$16$"$ &   1.8 $\pm$ 0.5 &  96.8 & 38.70 & \\
NGC 5055 X-7  &         & 13$^h$15$^m$49.9$^s$ & 42$^{\circ}$01$'$38$"$ &   3.2 $\pm$ 0.7 &   9.9 & 38.94 & N \\
NGC 5055 X-8  &         & 13$^h$15$^m$51.7$^s$ & 42$^{\circ}$01$'$34$"$ &   2.1 $\pm$ 0.5 &  25.3 & 38.76 & \\
NGC 5055 X-9  &         & 13$^h$16$^m$02.6$^s$ & 42$^{\circ}$01$'$48$"$ &   1.6 $\pm$ 0.4 & 142.9 & 38.65 & \\
NGC 5194 X-1  &         & 13$^h$29$^m$40.1$^s$ & 47$^{\circ}$12$'$35$"$ &   2.6 $\pm$ 0.3 & 140.5 & 38.91 & \\
NGC 5194 X-2  &         & 13$^h$29$^m$43.9$^s$ & 47$^{\circ}$11$'$30$"$ &   1.4 $\pm$ 0.3 &  96.5 & 38.64 & \\
NGC 5194 X-3  &         & 13$^h$29$^m$46.5$^s$ & 47$^{\circ}$10$'$38$"$ &   1.4 $\pm$ 0.3 & 100.3 & 38.65 & \\
NGC 5194 X-4  &         & 13$^h$29$^m$53.2$^s$ & 47$^{\circ}$11$'$40$"$ &  28.4 $\pm$ 1.0 &  11.5 & 39.95 & N \\
NGC 5194 X-5  &         & 13$^h$29$^m$54.4$^s$ & 47$^{\circ}$14$'$33$"$ &   1.2 $\pm$ 0.2 & 161.7 & 38.57 & \\
NGC 5194 X-6  &         & 13$^h$29$^m$59.6$^s$ & 47$^{\circ}$15$'$48$"$ &   2.5 $\pm$ 0.4 & 245.6 & 38.91 & \\
NGC 5194 X-7  &         & 13$^h$30$^m$01.6$^s$ & 47$^{\circ}$13$'$40$"$ &   5.8 $\pm$ 0.5 & 138.2 & 39.27 & \\
NGC 5194 X-8  &         & 13$^h$30$^m$06.9$^s$ & 47$^{\circ}$08$'$30$"$ &   0.7 $\pm$ 0.2 & 245.6 & 38.33 & \\
NGC 5194 X-9  &         & 13$^h$30$^m$08.1$^s$ & 47$^{\circ}$11$'$01$"$ &   7.2 $\pm$ 0.5 & 160.5 & 39.36 & \\
NGC 5195 X-1  &         & 13$^h$30$^m$06.8$^s$ & 47$^{\circ}$15$'$39$"$ &   1.3 $\pm$ 0.2 &  80.7 & 38.77 & \\
NGC 5204 X-1  & AB      & 13$^h$29$^m$39.2$^s$ & 58$^{\circ}$25$'$01$"$ &  30.3 $\pm$ 1.1 &  19.6 & 39.57 & N \\
NGC 5273 X-1  &         & 13$^h$42$^m$08.3$^s$ & 35$^{\circ}$39$'$09$"$ &   2.3 $\pm$ 0.5 &  16.7 & 39.70 & N \\
NGC 5457 X-1  & CD      & 14$^h$02$^m$03.9$^s$ & 54$^{\circ}$18$'$26$"$ &   0.6 $\pm$ 0.1 & 642.2 & 37.93 & \\
NGC 5457 X-2  & ABCD    & 14$^h$02$^m$22.5$^s$ & 54$^{\circ}$17$'$52$"$ &   0.8 $\pm$ 0.1 & 500.0 & 38.06 & v \\
NGC 5457 X-3  & ABCD    & 14$^h$02$^m$28.6$^s$ & 54$^{\circ}$16$'$21$"$ &   1.8 $\pm$ 0.1 & 499.4 & 38.44 & \\
NGC 5457 X-4  & BCD     & 14$^h$02$^m$28.6$^s$ & 54$^{\circ}$12$'$37$"$ &   1.3 $\pm$ 0.1 & 656.8 & 38.30 & \\
NGC 5457 X-5  & C       & 14$^h$02$^m$29.0$^s$ & 54$^{\circ}$26$'$29$"$ &   0.6 $\pm$ 0.1 & 508.4 & 37.95 & \\
NGC 5457 X-6  & ABCD    & 14$^h$02$^m$30.3$^s$ & 54$^{\circ}$21$'$15$"$ &   3.7 $\pm$ 0.1 & 389.1 & 38.74 & \\
NGC 5457 X-7  & C       & 14$^h$02$^m$34.8$^s$ & 54$^{\circ}$21$'$11$"$ &   0.3 $\pm$ 0.1 & 349.0 & 37.61 & \\
NGC 5457 X-8  & B       & 14$^h$02$^m$47.2$^s$ & 54$^{\circ}$26$'$53$"$ &   1.0 $\pm$ 0.2 & 415.7 & 38.17 & \\
NGC 5457 X-9  & CD      & 14$^h$02$^m$53.3$^s$ & 54$^{\circ}$21$'$08$"$ &   0.6 $\pm$ 0.1 & 187.7 & 37.93 & \\
NGC 5457 X-10 & ABCD    & 14$^h$03$^m$04.4$^s$ & 54$^{\circ}$27$'$32$"$ &   3.2 $\pm$ 0.1 & 388.8 & 38.68 & v \\
NGC 5457 X-11 & ACD     & 14$^h$03$^m$13.3$^s$ & 54$^{\circ}$20$'$51$"$ &   0.9 $\pm$ 0.1 &  27.2 & 38.15 & \\
NGC 5457 X-12 & D       & 14$^h$03$^m$13.6$^s$ & 54$^{\circ}$20$'$02$"$ &   0.5 $\pm$ 0.1 &  72.8 & 37.89 & \\
NGC 5457 X-13 & B       & 14$^h$03$^m$16.0$^s$ & 54$^{\circ}$27$'$30$"$ &   0.7 $\pm$ 0.2 & 375.6 & 38.03 & \\
NGC 5457 X-14 & AB      & 14$^h$03$^m$20.4$^s$ & 54$^{\circ}$17$'$15$"$ &   1.7 $\pm$ 0.2 & 244.3 & 38.41 & \\
NGC 5457 X-15 & CD      & 14$^h$03$^m$22.4$^s$ & 54$^{\circ}$19$'$41$"$ &   0.8 $\pm$ 0.1 & 115.0 & 38.06 & \\
NGC 5457 X-16 & BC      & 14$^h$03$^m$24.0$^s$ & 54$^{\circ}$19$'$41$"$ &   0.6 $\pm$ 0.1 & 123.7 & 37.99 & \\
NGC 5457 X-17 & B       & 14$^h$03$^m$31.4$^s$ & 54$^{\circ}$21$'$55$"$ &   0.9 $\pm$ 0.2 & 151.1 & 38.14 & \\
NGC 5457 X-18 & D       & 14$^h$03$^m$32.9$^s$ & 54$^{\circ}$20$'$58$"$ &   1.1 $\pm$ 0.2 & 159.5 & 38.23 & \\
NGC 5457 X-19 & BCD     & 14$^h$03$^m$36.5$^s$ & 54$^{\circ}$19$'$21$"$ &   0.6 $\pm$ 0.1 & 221.3 & 37.95 & \\
NGC 5457 X-20 & D       & 14$^h$03$^m$41.8$^s$ & 54$^{\circ}$19$'$02$"$ &   0.5 $\pm$ 0.1 & 271.2 & 37.91 & \\
NGC 5457 X-21 & D       & 14$^h$03$^m$51.2$^s$ & 54$^{\circ}$24$'$09$"$ &   0.6 $\pm$ 0.1 & 362.9 & 37.93 & \\
NGC 5457 X-22 & BCD     & 14$^h$03$^m$54.6$^s$ & 54$^{\circ}$21$'$55$"$ &   0.9 $\pm$ 0.1 & 350.3 & 38.12 & \\
NGC 5457 X-23 & D       & 14$^h$04$^m$00.4$^s$ & 54$^{\circ}$09$'$09$"$ &   0.4 $\pm$ 0.1 & 828.4 & 37.75 & \\
NGC 5457 X-24 & ABCD    & 14$^h$04$^m$15.0$^s$ & 54$^{\circ}$26$'$02$"$ &   4.7 $\pm$ 0.1 & 599.9 & 38.84 & \\
NGC 5457 X-25 & CD      & 14$^h$04$^m$17.5$^s$ & 54$^{\circ}$16$'$10$"$ &   0.6 $\pm$ 0.1 & 627.7 & 37.92 & \\
NGC 5457 X-26 & ABCD    & 14$^h$04$^m$22.5$^s$ & 54$^{\circ}$19$'$20$"$ &   1.1 $\pm$ 0.1 & 603.2 & 38.21 & v \\
NGC 5457 X-27 & CD      & 14$^h$04$^m$29.7$^s$ & 54$^{\circ}$23$'$51$"$ &   0.7 $\pm$ 0.1 & 673.3 & 38.00 & \\
NGC 5905 X-1  &         & 15$^h$15$^m$23.7$^s$ & 55$^{\circ}$30$'$57$"$ &   1.0 $\pm$ 0.2 &   8.4 & 40.01 & N \\
NGC 6217 X-1  & ABCDE   & 16$^h$32$^m$40.6$^s$ & 78$^{\circ}$11$'$49$"$ &   2.8 $\pm$ 0.2 &   4.2 & 40.02 & N \\
NGC 6503 X-1  &         & 17$^h$49$^m$12.8$^s$ & 70$^{\circ}$09$'$24$"$ &   1.5 $\pm$ 0.4 &  82.8 & 38.57 & \\
NGC 6946 X-1  & AB      & 20$^h$34$^m$26.5$^s$ & 60$^{\circ}$09$'$06$"$ &   1.3 $\pm$ 0.2 & 189.4 & 38.63 & \\
NGC 6946 X-2  & A       & 20$^h$34$^m$35.0$^s$ & 60$^{\circ}$10$'$29$"$ &   0.5 $\pm$ 0.1 & 141.1 & 38.22 & \\
NGC 6946 X-3  & AB      & 20$^h$34$^m$37.0$^s$ & 60$^{\circ}$09$'$26$"$ &   2.7 $\pm$ 0.2 & 110.4 & 38.96 & \\
NGC 6946 X-4  & A       & 20$^h$34$^m$49.4$^s$ & 60$^{\circ}$05$'$48$"$ &   0.7 $\pm$ 0.2 & 218.3 & 38.38 & \\
NGC 6946 X-5  & A       & 20$^h$34$^m$50.1$^s$ & 60$^{\circ}$12$'$15$"$ &   0.6 $\pm$ 0.1 & 170.6 & 38.28 & \\
NGC 6946 X-6  & AB      & 20$^h$34$^m$53.1$^s$ & 60$^{\circ}$09$'$07$"$ &   2.6 $\pm$ 0.2 &  20.9 & 38.95 & N \\
NGC 6946 X-7  & A       & 20$^h$34$^m$57.4$^s$ & 60$^{\circ}$08$'$25$"$ &   0.7 $\pm$ 0.2 &  72.3 & 38.39 & \\
NGC 6946 X-8  & A       & 20$^h$34$^m$58.5$^s$ & 60$^{\circ}$09$'$37$"$ &   0.6 $\pm$ 0.2 &  50.9 & 38.33 & \\
NGC 6946 X-9  & AB      & 20$^h$35$^m$00.9$^s$ & 60$^{\circ}$09$'$03$"$ &   3.5 $\pm$ 0.2 &  71.2 & 39.07 & \\
NGC 6946 X-10 & AB      & 20$^h$35$^m$01.2$^s$ & 60$^{\circ}$10$'$06$"$ &   0.8 $\pm$ 0.1 &  80.9 & 38.44 & \\
NGC 6946 X-11 & AB      & 20$^h$35$^m$01.4$^s$ & 60$^{\circ}$11$'$27$"$ &  16.4 $\pm$ 0.5 & 141.4 & 39.75 & \\
NGC 6946 X-12 & A       & 20$^h$35$^m$12.9$^s$ & 60$^{\circ}$07$'$26$"$ &   0.6 $\pm$ 0.1 & 223.7 & 38.31 & \\
NGC 6946 X-13 & A       & 20$^h$35$^m$19.4$^s$ & 60$^{\circ}$10$'$53$"$ &   0.6 $\pm$ 0.1 & 378.3 & 38.29 & \\
NGC 7331 X-1  &         & 22$^h$37$^m$04.4$^s$ & 34$^{\circ}$24$'$51$"$ &   2.8 $\pm$ 0.4 &  18.0 & 39.65 & N \\\hline
\end{tabular}
\end{table*}

\section{UPPER LIMITS ON NUCLEAR X-RAY LUMINOSITY}

95\% upper limits on the nuclear X-ray luminosities were
derived for those nuclei without a detected nuclear X-ray source. The results
are listed in Table~\ref{apptab2}.  Where there are two or more observations
of a source, the observation with the longest exposure was employed.

\begin{table}
\centering
\caption{95\% upper limits on the nuclear X-ray luminosity for those galaxies
with no detected nuclear source.}
\label{apptab2}
\begin{tabular}{lcc}\hline
Galaxy	& Count rate	& log \lx \\
	& ($\times 10^{-3} \ctsec$) 	& ($\ergsec$) \\\hline
IC 10    &  0.38 & 37.05 \\
NGC 147  &  0.63 & 36.43 \\
NGC 185  &  0.74 & 36.51 \\
NGC 205  &  1.24 & 36.66 \\
NGC 520  &  1.44 & 39.84 \\
NGC 891  &  1.04 & 38.86 \\
NGC 1058 &  0.36 & 38.33 \\
NGC 1560 &  0.86 & 37.84 \\
NGC 2276 &  1.72 & 40.23 \\
NGC 2366 &  0.68 & 37.57 \\
NGC 2403 &  0.30 & 37.55 \\
NGC 2775 &  1.55 & 39.48 \\
NGC 2976 &  0.70 & 37.32 \\
NGC 3073 &  0.77 & 39.14 \\
NGC 3185 &  0.68 & 39.25 \\
NGC 3193 &  1.47 & 39.66 \\
NGC 3294 &  1.76 & 39.83 \\
NGC 3310 &  1.25 & 39.34 \\
NGC 3384 &  1.45 & 38.76 \\
NGC 3389 &  1.08 & 39.52 \\
NGC 3395 &  0.74 & 39.49 \\
NGC 3628 &  1.15 & 38.58 \\
NGC 4214 &  1.15 & 37.89 \\
NGC 4395 &  2.46 & 38.22 \\
NGC 4470 &  1.17 & 39.80 \\
NGC 4527 &  2.53 & 39.41 \\
NGC 4631 &  1.30 & 38.51 \\
NGC 4638 &  1.40 & 39.36 \\
NGC 4647 &  1.60 & 39.42 \\
NGC 4656 &  1.35 & 38.56 \\
NGC 4772 &  2.24 & 39.52 \\
NGC 5195 &  2.11 & 38.99 \\
NGC 5354 &  1.43 & 39.89 \\
NGC 5457 &  0.48 & 37.85 \\
NGC 5775 &  0.66 & 39.47 \\
NGC 5850 &  1.15 & 39.79 \\
NGC 6503 &  2.96 & 38.86 \\
NGC 6654 &  1.05 & 39.81 \\\hline
\end{tabular}
\end{table}

\end{document}